\documentclass{aa}  

\usepackage{graphicx}
\usepackage{xcolor}
\usepackage{txfonts}
\usepackage{booktabs}
\usepackage{mathrsfs}
\newcommand{\dd}[1]{\,\text{d}#1}

\begin{document} 

    \title{Cosmology with the submillimetre galaxies magnification bias:\\ Proof of concept.}
    \titlerunning{Cosmology with magnification bias}
    \authorrunning{Bonavera L. et al.}

    \author{Bonavera L.\inst{1,2}, Gonz{\'a}lez-Nuevo J.\inst{1,2}, Cueli M. M.\inst{1,2}, Ronconi T.\inst{3,4,5}, Migliaccio M.\inst{6,7}, Dunne L.\inst{8}, Lapi A.\inst{3,5}, Maddox S.J.\inst{8}, Negrello M.\inst{8}}

   \institute{$^1$Departamento de Fisica, Universidad de Oviedo, C. Federico Garcia Lorca 18, 33007 Oviedo, Spain\\
             $^2$Instituto Universitario de Ciencias y Tecnologias Espaciales de Asturias (ICTEA), C. Independencia 13, 33004 Oviedo, Spain\\
             $^3$International School for Advanced Studies (SISSA), via Bonomea 265, I-34136 Trieste, Italy\\
             $^4$INFN Sezione di Trieste, via Valerio 2, I-34127 Trieste, Italy\\
             $^5$Institute for Fundamental Physics of the Universe (IFPU), Via Beirut 2, I-34014 Trieste, Italy\\
             $^6$Dipartimento di Fisica, Universit\`a di Roma Tor Vergata, Via della Ricerca Scientifica, 1, Roma, Italy\\
             $^7$INFN, Sezione di Roma 2, Universit\`a di Roma Tor Vergata, Via della
Ricerca Scientifica, 1, Roma, Italy\\
             $^8$School of Physics \& Astronomy, Cardiff University, The Parade, Cardiff CF24 3AA, UK\\
             }

   \date{Received xxx, xxxx; accepted xxx, xxxx}

 
  \abstract
   {As recently demonstrated,   high-z submillimetre galaxies (SMGs) are the perfect background sample for tracing the mass density profiles of galaxies and clusters (baryonic and dark matter) and their time-evolution through gravitational lensing. Their magnification bias, a weak gravitational lensing effect, is a powerful tool for constraining the free parameters of a halo occupation distribution (HOD) model and potentially also some of the main cosmological parameters.}
   {The aim of this work is to test the capability of the magnification bias produced on high-z SMGs as a cosmological probe. We exploit cross-correlation data to constrain not only astrophysical parameters ($M_{min}$, $M_1$, and $\alpha$), but also some of the cosmological ones ($\Omega_m$, $\sigma_8$, and $H_0$) for this proof of concept.}
   {The measured cross-correlation function between a foreground sample of GAMA galaxies with spectroscopic redshifts in the range 0.2 < z < 0.8 and a background sample of H-ATLAS galaxies with photometric redshifts >1.2 is modelled using the traditional halo model description that depends on   HOD and   cosmological parameters. These parameters are then estimated by performing a Markov chain Monte Carlo analysis using different sets of priors to test the robustness of the results and to study the performance of this novel observable with the current set of data.}
   {With our current results, $\Omega_m$ and $H_0$ cannot be well constrained. However, we can set a lower limit of >0.24 at 95\% confidence level (CL) on $\Omega_m$ and we see a slight trend towards $H_0>70$ values. For our constraints on $\sigma_8$ we obtain only a tentative peak around 0.75, but an interesting upper limit of $\sigma_8\lesssim 1$ at 95\% CL. We also study the possibility to derive better constraints by imposing more restrictive priors on the astrophysical parameters.}
   {}


   \maketitle
%

\section{Introduction}

Magnification bias is the apparent excess number of high redshift sources seen near low redshift structures, and it can be quantified using the cross-correlation between high and low redshift galaxies.  Light rays coming from a distant source are deflected by the intervening gravitational field so that the apparent area of a given region of sky is stretched. The increased area decreases the apparent surface density of any background sources.  The deflections also amplify the flux of the background sources, increasing their chances of being included in a flux-limited sample. The net effect is known as magnification bias, and it has been extensively described in the literature \citep[see e.g.][]{SCH92}.

The cross-correlation between two source samples with non-overlapping redshift distributions is an unambiguous manifestation of magnification bias. The occurrence of such correlations has been tested and established in several contexts: an $8\sigma$ detection of cosmic magnification from the galaxy-quasar cross-correlation function \citep{SCR05}; a simultaneous detection of gravitational magnification and dust reddening effects due to galactic halos and large-scale structure \citep[galaxy-quasar cross-correlation;][]{MEN10}; and a $7\sigma$ detection of a cross-correlation signal between $z\sim3-5$ Lyman-break galaxies and Herschel sources \citep[][]{HIL13}, among others.

The strength of the magnification bias  depends on the size of gravitational deflection as light passes near low redshift galaxies, and since the deflection depends on the cosmological distances, and on the galaxy halo properties, we can use these measurements to set constraints on cosmological parameters.  The aim of our current analysis is to investigate how our measurements of magnification bias can constrain parameters in the standard cosmological model.

The current standard cosmological model successfully accounts for anisotropies in the CMB \citep[e.g.][]{HIN13, PLA16_XIII, PLA18_VI}, the accelerating expansion of the Universe observed from SNIa \cite[e.g.][]{BET14}, and Big Bang nucleosynthesis \citep[e.g.][]{FIE06}, and also predicts some important features of large-scale structure (LSS) in the distribution of galaxies (e.g. \citealt{PEA01}) including baryon acoustic oscillations (BAOs; e.g. \citealt{ROS15}). This means that independent and complementary constraints on the model parameters can be obtained by measuring such observables \citep[e.g.][]{PEA94}. The consistency between different probes is a key reason that the model is considered to be so successful.  However, with the increase in the quality and quantity of the measurements, some tensions and small-scale issues have arisen, such as the missing satellite problem \citep{MOO99, KLY99} and the cusp-core problem \citep{WEC02}, that might indicate the necessity of modifications of the $\Lambda$CDM model.

One of the main tensions is in the value of the Hubble constant, $H_0$: local estimates suggest a relatively high value \citep[$74.03 \pm 1.42$ km/s/Mpc,][]{RIE19}, whereas CMB and BAOs analysis advise a relatively low value \citep[$67.4 \pm 0.5$ km/s/Mpc][]{PLA18_VI}. Another important inconsistency lies in the usually degenerate relationship between the $\Omega_m$ and $\sigma_8$ parameters. Constraints from local observables, such as LSS in the galaxy distribution, tend to prefer lower values compared to those derived from high redshift observables such as the CMB \citep[e.g.][]{HEY13,PLA16_XXIV,HIL17}. This $\sim 2\sigma$ tension is discussed in \cite{PLA18_VI}.  With this in mind, any method that constrains cosmological parameters using a new set of observables is worth pursuing in the effort to resolve the tensions. At the moment our constraints from the magnification bias are relatively weak, but the use of new independent observables makes it a valuable new technique.

In this paper we use the cross-correlation function between low and high redshift galaxy samples to measure magnification bias.  An optimal choice of foreground and background samples can enhance the cross-correlation signal, and in particular some features of the  submillimetre galaxies (SMGs) make them close to the optimal background sample for lensing studies. Some of their most important characteristics, related to their exploitation in magnification bias studies, are summarised below.

Firstly, data collected before the advent of the European Herschel Space Observatory \citep[Herschel;][]{PIL10} and the South Pole Telescope \citep[SPT,][]{CAR11} suggested that the number density of SMGs drops off abruptly at relatively bright submillimetre flux densities ($\sim 50 mJy$ at $500 \mu m$), indicating a steep luminosity function and a strong cosmic evolution for this class of sources \citep[e.g. ][]{GRA04}. Several authors have argued that the bright tail of the submillimetre number counts may contain a significant fraction of strongly lensed SMGs \citep{BLA96, NEG07}. The Herschel Multi-tiered Extragalactic Survey \citep[HerMES;][]{OLI10} and the Herschel Astrophysical Terahertz Large Area Survey \citep[H-ATLAS;][]{EAL10} are wide-field surveys ($\sim 380 deg^2$ and $\sim 550 deg^2$, respectively) conducted by the Herschel satellite. Thanks to their sensitivity and frequency coverage both surveys have led to the discovery of several lensed SMGs \citep[][and references therein]{NEG17}. The selection of strongly lensed galaxies at these wavelengths is made possible by the (predicted) steep number counts of SMGs \citep{BLA96, NEG07}; in fact, almost all of those galaxies whose flux density has been boosted by an event of lensing can be observed above a certain threshold, namely $\sim 100 mJy$ at $500 \mu m$ \citep[see][]{NEG10, WAR13}.

Next, with Heschel data, \cite{NEG10} produced the first sample of five strongly lensed galaxies (SLGs) by means of a simple selection in flux density at 500 $\mu m$. Preliminary source catalogues derived from the full H-ATLAS were then used to identify the submillimetre brightest candidate lensed galaxies for follow-up observations with both ground-based and space telescopes to measure their redshifts \citep[see ][and references therein]{NEG17} with 80 SLG candidates  and confirm their nature \citep{NEG10,BUS12,BUS13,FU12,CAL14}.

Using the same methodology, i.e. a cut in flux density at 500 $\mu m$, \cite{WAR13} have identified 11 SLGs over 95 $deg^2$ of HerMES \citep{OLI10}, while more recently \cite{NAY16} have published a catalogue of 77 galaxy candidates at lenses with $F_{500} \ge 100$ mJy extracted from the HerMES Large Mode Survey \citep[HeLMS;][]{OLI12} and the Herschel Stripe 82 Survey \citep[HerS;][]{VIE14} over an area of 372 $deg^2$. Altogether, the extragalactic surveys carried out with Herschel delivered a sample of more than a hundred   submillimetre bright SLGs, which, as argued by \citet{GON12, GON19}, might increase to over a thousand if the selection is based on the steepness of the luminosity function of SMGs \citep{LAP11} rather than that of the number counts. Similarly, at millimetre wavelengths, the SPT survey has already discovered several tens of lensed galaxies \citep[e.g.][]{VIE13, SPI16} and other lensing events have been found in the \textit{Planck} all-sky surveys \citep{CAN15, HAR16}.

The properties of SMGs have been very well quantified by these previous analyses, and SMGs can be considered solid targets to be studied with lensing. In particular, the (sub)millimetre SLGs are very faint in the optical, whereas most foreground lenses are passive ellipticals \citep{AUG09, NEG10}, essentially invisible at submillimetre wavelengths. This means that the foreground lens is `transparent' at (sub)millimetre wavelengths, i.e. it does not confuse the measurements of the background source and vice versa. In this way, the (sub)millimetre selection shares with spectroscopic searches the capability of detecting lensing events with small impact parameters, with the added advantage that in most cases there is no need to subtract the lens contribution to recover the source images within the effective radii of the lenses. A further advantage given by (sub)millimetre selection is that the typical source redshifts are above $ z > 1-1.5 $, and so the samples probe earlier phases of galaxy evolution, which  typically have higher lensing optical depths. This makes the SMGs the perfect background sample for tracing the mass density profiles of galaxies and clusters, including baryonic and dark matter, and their time-evolution, using gravitational lensing.

A first attempt at measuring lensing-induced cross-correlations between Herschel/SPIRE galaxies and low-z galaxies was carried out by \cite{WAN11}, who found convincing evidence of the effect. With much better statistics, this possible bias was studied in detail in \cite{GON14} by measuring the angular cross-correlation function between selected H-ATLAS high-z sources, $z > 1.5$, and two optical samples with redshifts $0.2 < z < 0.8$, extracted from the Sloan Digital Sky Survey \citep[SDSS;][]{AHN12} and Galaxy and Mass Assembly \citep[GAMA,][]{DRI11} surveys. The observed cross-correlation function was measured with high significance, $> 10\sigma$. Moreover, based on realistic simulations, it was concluded that the signal was entirely explained by a magnification bias produced by the weak lensing effect of low redshift cosmic structures (galaxy groups and clusters with halo masses in the range $10^{13.2}-10^{14.5}M_{\odot}$) signposted by the brightest galaxies in the optical samples.

\cite{GON17} used updated catalogues covering a wider area to make more accurate measurements of the cross-correlation function, with S/N > 5 above 10 arcmin and reaching S/N$\sim$20 below 30 arcsec.  Thanks to the better statistics it was possible to split the sample into different redshift bins and to perform a tomographic analysis (with S/N > 3 above 10 arcmin and reaching S/N$\sim$15 below 30 arcsec). Moreover, a halo model was implemented to extract astrophysical information about the background galaxies and the deflectors that produce the lensing that links the foreground (lenses) and background (sources) samples. In the case of the sources,   typical mass values were found to be in agreement with previous studies \citep[$10^{12.1}-10^{12.9} M_{\odot}$, ][]{Xia12, COO10, MIT12, WIL17}. However, the lenses are massive galaxies or even galaxy groups and clusters, with minimum mass of $M_{lens}\sim10^{13.06}M_{\odot}$, confirming the interpretation from \cite{GON14}. The halo modelling also helped to identify a strong lensing contribution to the cross-correlation for angular scales below 30 arcsec. This interpretation is supported by the results of basic but effective simulations.

Following on from this, \cite{BON19} were able to measure the magnification bias produced by a sample of quasi-stellar objects QSOs acting as lenses, $0.2<z<1.0$, on the SMGs observed by Herschel at $1.2<z<4.0$, by applying the traditional cross-correlation function approach and the Davis-Peebles estimator through a stacking technique. Using the halo modelling of the cross-correlation function it was possible to estimate the halo mass where the QSOs acting as lenses are located: it was found to be greater than $M_{min} > 10^{13.6_{-0.4}^{+0.9}} M_{\odot}$, also confirmed by the mass density profile analysis ($M_{200c}\sim 10^{14} M_{\odot}$). These mass values indicate that we are observing the lensing effect of a cluster size halo signposted by the QSOs.

In the rest of this paper, we describe our first attempts to estimate cosmological parameters using the cross-correlation function and data from  \cite{GON17}, and demonstrate the powerful potential of magnification bias as a cosmological tool. In particular, we focus on the estimation of the parameters $\Omega_m$, $\sigma_8$, and $H_0$   in order to quantify what can actually be done with magnification bias, and what will be needed in the future to improve our cosmological analysis.

The work  is organised as follows. In section \ref{sec:data} the background and foreground samples are described and in section \ref{sec:methodology} the methodology is presented, both for auto- and cross-correlation measurements. Our results and conclusions are discussed in sections \ref{sec:results} and \ref{sec:conclusion}, respectively. In Appendix \ref{appendix:sens_plots} we show how the variations in the astrophysical and cosmological parameters  affect the model behaviour and we compare the model with the data.

\section{Data}
\label{sec:data}

\subsection{Background sample}
\label{sec:bk_sample}

The background sample was selected from the H-ATLAS data set, the largest  extra-galactic survey carried out by the Herschel space observatory \citep{PIL10}: its two instruments the Photodetector Array Camera and Spectrometer \citep[PACS, ][]{POG10} and the Spectral and Photometric Imaging REceiver \citep[SPIRE, ][]{GRI10} cover about 610 $deg^2$ between 100 and 500 $\mu m$. The survey consists of five different fields: three located on the celestial equator \citep[GAMA fields or Data Delivery 1 (DR1);][]{VAL16, BOU16, RIG11, PAS11, IBA10} covering a total area of 161.6 $deg^2$ and two centred on the north and south galactic poles \citep[NGP and SGP fields or DR2;][]{SMI17, MAD18} covering 180.1 $deg^2$ and 317.6 $deg^2$, respectively. 

\begin{figure}[ht]
\centering
\includegraphics[width=9cm]{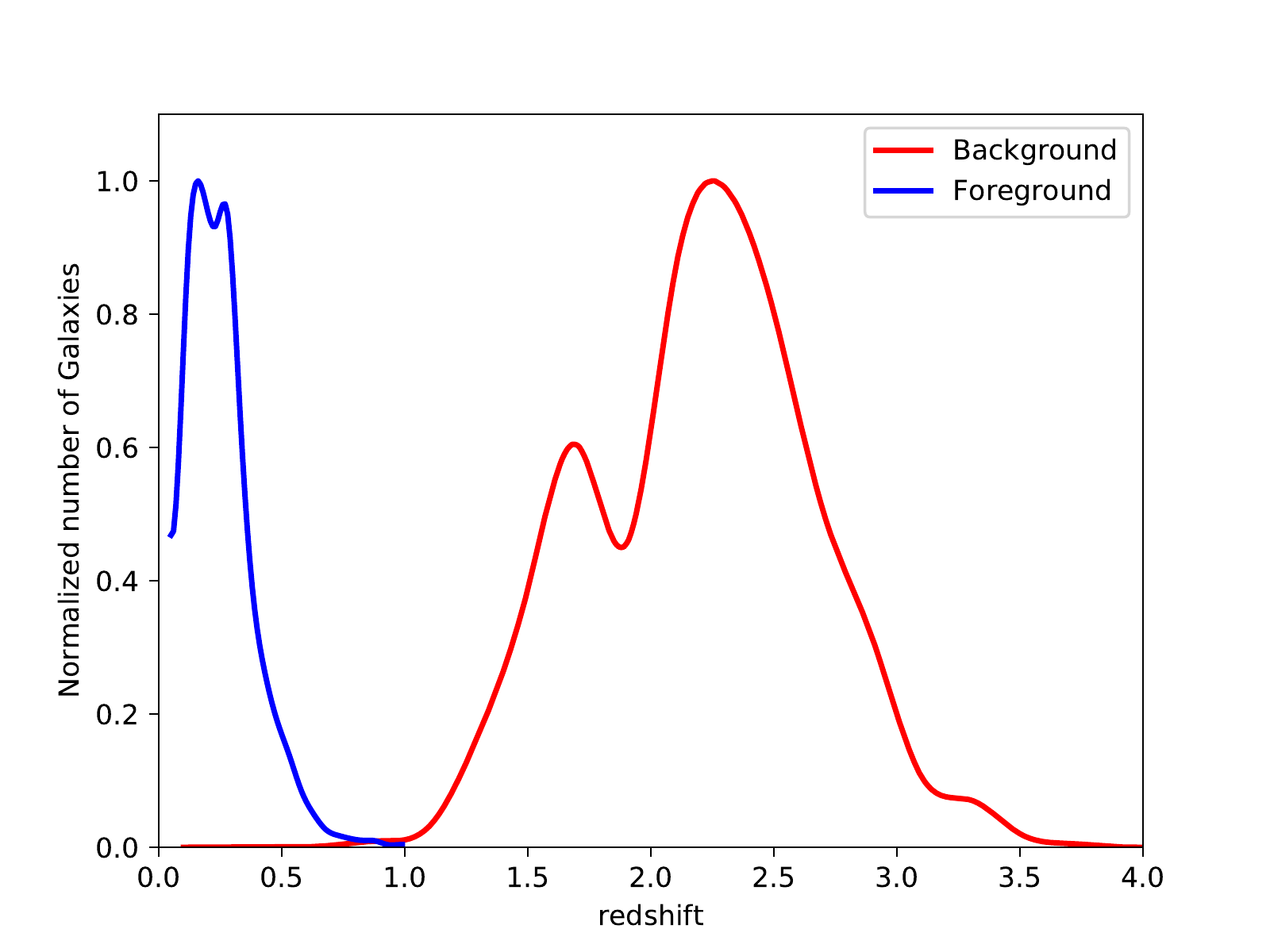}
\caption{Redshift distributions of the background H-ATLAS sample (red) and the foreground GAMA sample (blue).}
\label{Fig:dNdz_data}
\end{figure}

We selected our background sample from the sources detected in the three GAMA fields, covering a total area of $\sim 147 deg^2$, and the part of the SGP that overlaps with the foreground sample ($\sim 60 deg^2$, see section \ref{sec:fore_sample}). This leads to $\sim 207 deg^2$ of common area between the two samples.

It should be noted that an implicit $4\sigma$ detection limit at $250 \mu m$ ($\sim S250 > 29$ mJy) was applied to both H-ATLAS DRs. At this frequency the estimated $1\sigma$ source detection total noise level (both instrumental and confusion) is 7.4 mJy for both DR1 and DR2 catalogues \citep{VAL16, MAD18}. In addition, a $3\sigma$ limit at $350 \mu m$ was introduced \citep{GON17} to increase the robustness of the photometric redshift estimation.

In order to avoid overlap in the redshift distributions of lenses and background sources, only background sources with photometric redshift 1.2 < z < 4.0 were   taken into account. 
Such redshifts were estimated with a minimum $\chi^2$ fit of a template SED to the SPIRE data (using PACS data when possible). The adopted template is the SED of SMM J2135-0102 \citep[`the Cosmic Eyelash' at z = 2.3;][]{IVI10,SWI10}, proven to be the best overall template with $\Delta z/(1 + z) = -0.07$ and dispersion of 0.153 \citep{IVI16,GON12,LAP11}. In the end, we are left with 57930 sources that correspond approximately to  24\%\   of the initial sample.

Figure \ref{Fig:dNdz_data} shows the redshift distribution of the background sample in red, being $\langle z \rangle = 2.2^{+0.4}_{-0.5}$ the mean redshift of the sample (the uncertainty indicates the $1\sigma$ limits). This redshift distribution is the estimated $p(z|W)$ of the galaxies selected by our window function (i.e. a top hat for 1.2 < z < 4.0). It takes into account the effect of random errors in photometric redshifts on the redshift distribution, as  done in \cite{GON17}.

\subsection{Foreground sample}
\label{sec:fore_sample}

As detailed  in \cite{GON17}, the foreground sources were drawn from the GAMA II \citep{DRI11, BAL10,BAL14,LIS15} spectroscopic survey. Moreover, both H-ATLAS and GAMA II surveys were coordinated in order to maximise the common area coverage overlap. In particular, the three equatorial regions at 9, 12, and 14.5 h (referred to as G09, G12, and G15, respectively) were observed by both surveys, and the SGP region surveyed by H-ATLAS was partially covered by GAMA II. We are then left with a common area of  about $\sim 207 deg^2$  between the two surveys, which has been surveyed down to a limit of $r \simeq 19.8$ mag.

Our main sample consists of a selection of all GAMA II galaxies with reliable redshift measurements between 0.2 < z < 0.8. This main sample comprises a total of $\sim 150000$ galaxies, whose median redshift, $z_{spec,med} = 0.28$, is significantly lower than that of the background sample, as shown by the blue histogram in Figure \ref{Fig:dNdz_data}.

The lower limit was chosen because we do not expect a relevant gravitational lensing effect from mater density distribution at z < 0.2 for the background sample \citep[see e.g.][]{LAP12}. For a similar reason, we can safely neglect any possible magnification bias affecting the lens sample: the lensing optical depth declines steeply for sources below z<0.5.

\section{Methodology}
\label{sec:methodology}

\subsection{Halo occupation distribution (HOD)}
Nowadays the power spectrum of the galaxy distribution is parametrised by the sum of a two-halo term, related to the correlation between different halos that dominates on large scales, and a one-halo term, more important on small scales, that depends on the distribution of galaxies within the same halo. 

All halos above a minimum mass $M_{min}$ host a galaxy at their centre, while any remaining galaxy is classified as satellite and such satellites are distributed proportionally to the halo mass profile \citep[see e.g.][] {ZHE05}. Halos host satellites when their mass exceeds the $M_1$ mass, and the number of satellites is a power-law ($\alpha$) function of halo mass.

Therefore, the central galaxy occupation probability can be parametrised as a step function 
\begin{equation}
    \label{eq:ncen01}
    N_\text{cen}(M_h) =
    \begin{cases}
    0 \quad \text{if}\ M_h < M_\text{min}\\
    1 \quad \text{otherwise}
    \end{cases}
\end{equation}
and the satellite galaxies occupation as
\begin{equation}
    \label{eq:nsat01}
    N_\text{sat}(M_h) = N_\text{cen}(M_h) \cdot \biggl(\dfrac{M_h}{M_1}\biggr)^{\alpha_{\text{sat}}}
,\end{equation}
where $M_\text{min}$, $M_1$, and $\alpha_\text{sat}$ are the free-parameters of the model.

\subsection{Foreground angular auto-correlation function}
\label{sec:auto_corr}

In order to have an independent estimation of the HOD parameters that correspond to the foreground galaxies acting as lenses, we perform an auto-correlation analysis on the GAMA sample.

\subsubsection{Estimation methodology}

For this auto-correlation analysis, we focus on the three  largest fields:  G09, G12, and G15. They constitute $\sim63$\% of the whole GAMA survey and they are the only fields with complete overlap with the H-ATLAS fields.  We expect no relevant differences in their correlation properties compared to the rest of the survey.
Moreover, we perform a further reduction in the foreground sample by selecting the normalising redshift quality (catalogue column labelled `\texttt{NQ}') greater than $2$ and the spectroscopic redshift (catalogue column labelled `\texttt{Z}') in the range $0.05 < z_\text{spec} < 0.6$.
The resulting median redshift of the sample of  sources  is $z_\text{med} = 0.23$.

We estimate the number density of sources by computing the volume of each field as $V_\text{field} = A_\text{field} \cdot \Delta d_C$, where $\Delta d_C = d_C(z_\text{max}) - d_C(z_\text{min})$ is the difference in comoving distance between the maximum and minimum redshift selected.
The area of the field is computed as $A_\text{field} = d_\text{hav}(RA) \cdot d_\text{hav}(Dec),$ where $d_\text{hav}(RA)$ and $d_\text{hav}(Dec)$ are the haversine distances
\begin{equation}
    \label{eq:hav_dist}
     d_\text{hav} / r = \arcsin\biggl( \sqrt{ \sin\bigl(\dfrac{\phi_2 - \phi_1}{2}\bigr) + \cos \phi_1 \cos \phi_2 \bigl( \dfrac{\lambda_2 - \lambda_1}{2} \bigr)} \biggr)
\end{equation}
along the RA and Dec directions, respectively.

In Table~\ref{tab:ng_mes} we show the measures obtained for the three different selected fields with the cuts described above.
The last row of the table shows the median value of $n_g$ and maximum error value, $\sigma_{n_g}$, among the three fields.

\begin{table}[ht]
    \centering
    \begin{tabular}{ccc}
        \toprule
        Field & $n_g [$ Mpc$^{-3} ]$ & $\sigma_{n_g} [ \text{Mpc}^{-3} ]$ \\ \midrule
        \texttt{G09} & 1.528e-3 & 6.095e-6 \\
        \texttt{G12} & 1.052e-3 & 4.023e-6 \\
        \texttt{G15} & 1.180e-3 & 4.601e-6 \\
        \midrule
        \texttt{med} & 1.217e-3 & 6.095e-6 \\\noalign{\smallskip}
        \hline \bottomrule
    \end{tabular}
    \caption{Number density of sources in the three selected GAMA fields, with the cuts described in the text.}
    \label{tab:ng_mes}
\end{table}

\begin{figure}[ht]
    \centering
    \includegraphics[width=0.45\textwidth]{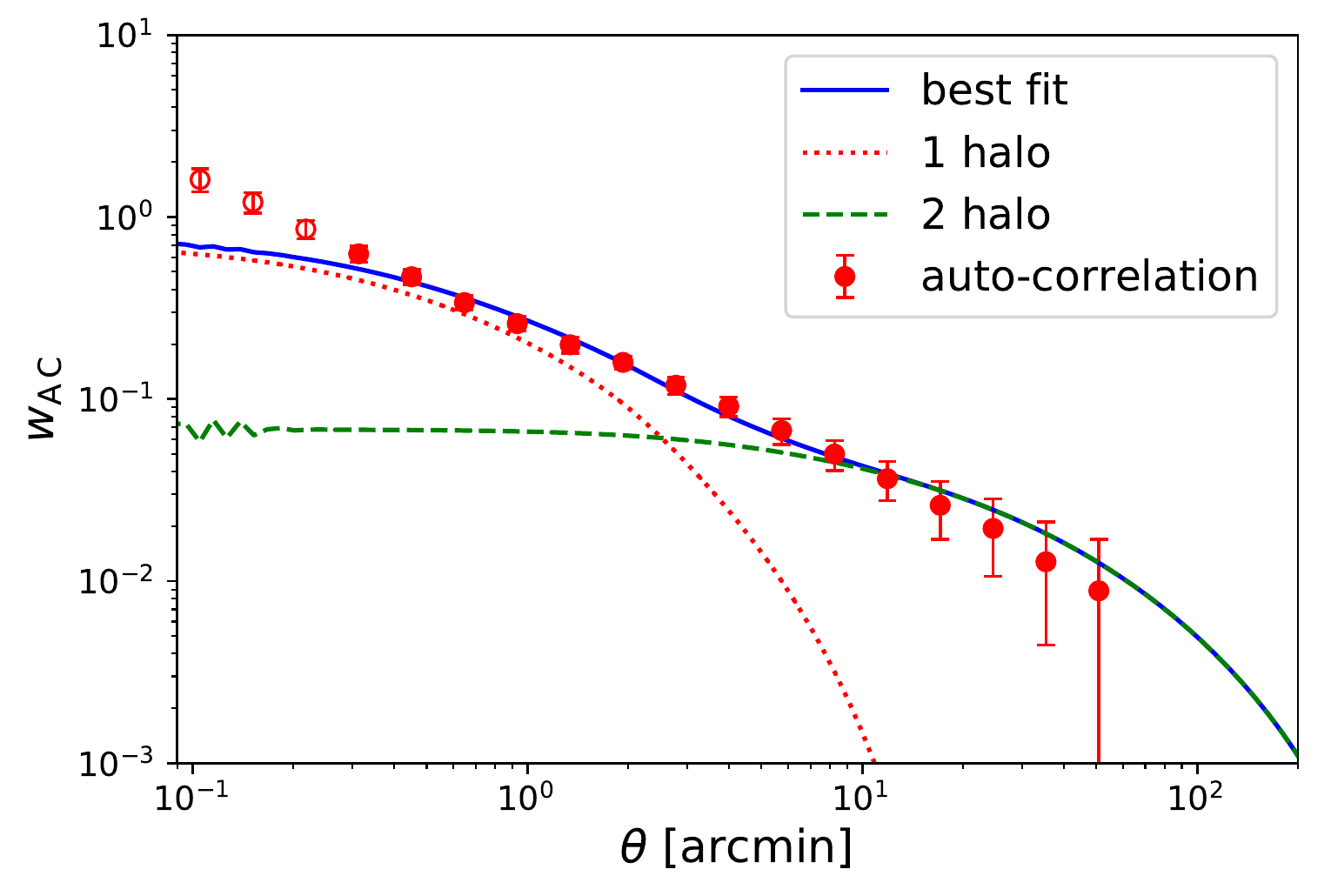}
    \caption{Measured angular auto-correlation function of the foreground sample (red circles). The best-fit halo model prediction is shown as a comparison (total, blue solid line; one-halo term, red dotted line; two-halo term, green dashed line).}
    \label{fig:acf_model}
\end{figure}

We compute the angular correlation function $\omega(\theta)$ in each field by applying the modified Landy-Szalay estimator \citep{LAN93}
\begin{equation}
    \label{eq:LS_est}
    w(\theta) = \dfrac{DD(\theta) - 2 DR(\theta) + RR(\theta)}{RR(\theta)}
\end{equation}
and estimate the error on this measure by dividing each field into 128 bootstrap samples (see Figure \ref{fig:acf_model}).

\subsubsection{Model}
\label{sec:auto_model}
The traditional halo modelling  \citep[see e.g.][]{COO02} allows us to obtain estimates of the average surface density of sources and of the angular correlation function at a given redshift.
The surface density is given by
\begin{equation}
    \label{eq:ng01}
    n_g(z) \equiv \int_{M_\text{min}}^{M_\text{max}} \langle N_g \rangle (M_h) n(M_h, z) \dd{M_h}
,\end{equation}
where $N_g(M_h) = N_\text{cen}(M_h) + N_\text{sat}(M_h)$ and where $n(M_h, z)$ is the Sheth \& Tormen mass function \citep{SHE99}.
The angular correlation function is given by
\begin{equation}
    \label{eq:wt01}
    \omega ( \theta, z ) = \int \dd{z} \dfrac{\dd{V}(z)}{\dd{z}} \mathscr{N}^2(z)\, w[ r_p(\theta, z), z ]
,\end{equation}
where $w[ r_p, z ]$ is the projected two-point correlation function obtained using the Limber approximation, $\frac{\dd{V}(z)}{\dd{z}}$ is the comoving volume unit, and $\mathscr{N}(z)$ is the normalised redshift distribution of the target population.  If we assume that $w(r_p, z)$ is approximately constant in the redshift interval $[z_1, z_2]$, we can then write
\begin{equation}
    \label{eq:wt02}
    \omega ( \theta, z ) \approx  \biggl[ \int_{z_1}^{z_2} \dd{z} \dfrac{\dd{V}(z)}{\dd{z}} \mathscr{N}^2(z) \biggr] \cdot w[ r_p( \theta ), \overline{z} ) ] = A \cdot w[ r_p( \theta ), \overline{z} ) ]
\end{equation}
so that the normalisation factor, $A$, can be accounted for by adding another parameter to the model.

\subsection{Angular cross-correlation function}
\label{sec:cross_corr}

\begin{figure}[ht]
\centering
\includegraphics[width=9cm]{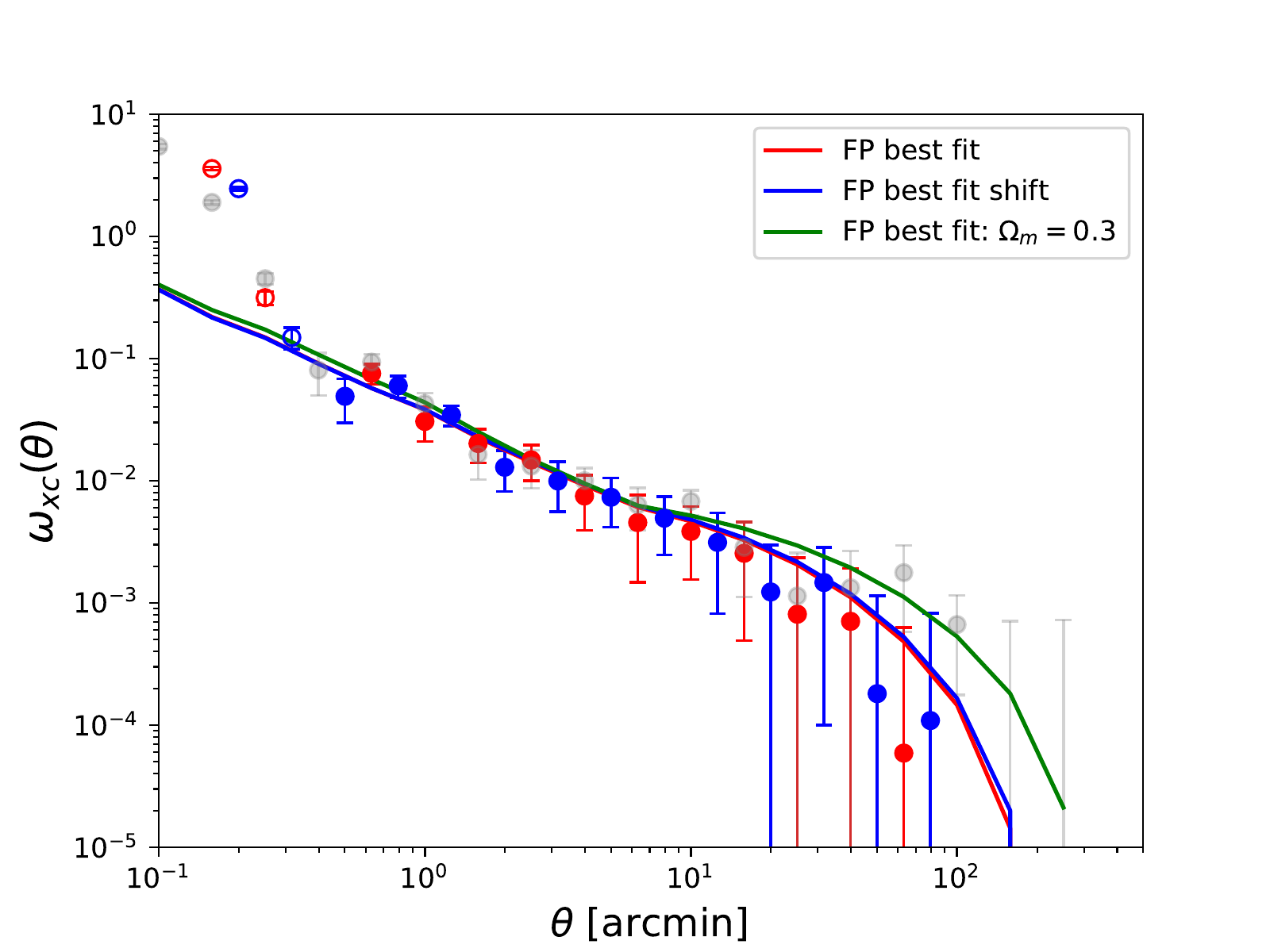}
    \caption{Cross-correlation data computed on Herschel SMGs (as background sample) and GAMA galaxies (as lenses sample). The grey points are exactly the results obtained in \cite{GON17}. The red data set \citep[obtained using the same binning as in][]{GON17} and the blue data set \citep[obtained using same the binning, but half-size bin shifted with respect to][]{GON17} are the cross-correlation results computed for this work, exploiting a larger area. The solid lines represent the best fit with flat priors to the astrophysical and cosmological parameters using the default binning (red) and the shifted binning (blue). The green line is the best fit to the red case, but adopting $\Omega_m=0.3$.}
    \label{Fig:xcorr_data}
\end{figure}

For the cross-correlation measurements we use the data described in section \ref{sec:data} and applied the same approach as in \cite{GON17} with slight changes, as described in the following section.

\subsubsection{Estimation method}
We use a modified version of the \cite{LAN93} estimator \citep{HER01},
\begin{equation}
w_x(\theta)=\frac{\rm{D}_1\rm{D}_2-\rm{D}_1\rm{R}_2-\rm{D}_2\rm{R}_1+\rm{R}_1\rm{R}_2}{\rm{R}_1\rm{R}_2}
,\end{equation}
where $\rm{D}_1\rm{D}_2$, $\rm{D}_1\rm{R}_2$, $\rm{D}_2\rm{R}_1$, and $\rm{R}_1\rm{R}_2$ are the normalised data1-data2, data1-random2, data2-random1, and random1-random2 pair counts for a given separation $\theta$.

In this work we do not apply exactly the same procedure used by \citet{GON17} and \citet{BON19}.  Instead, we follow the original estimation procedure of \citet{GON14}, which divides the total common area between the two samples into several square mini-regions with minimal overlap. Therefore, the estimated cross-correlation function in each mini-region can be assumed to be  statistically independent. In this way we can use the full available area (maximising the data statistics) without leaving portions of the mini-areas with no sources, as would be the case for the circular regions used by \citet{GON17}.  Taking into account the shape of the H-ATLAS areas (related with the satellite scanning strategy), we adopt square mini-regions of $\sim 4 deg^2$. This division provided us with 62 usable almost independent mini-regions (15 in each GAMA field and 17 in the SGP field). We verified that using bigger square mini-regions ($\sim 16 deg^2$) leads to results that are compatible within the uncertainties on large scales, but with higher uncertainties at intermediate scales due to the higher cosmic variance effect with only 16 independent mini-regions.

The estimated cross-correlation function is shown in Figure \ref{Fig:xcorr_data} (red circles). Each measurement corresponds to the mean value of the cross-correlation functions estimated in each individual mini-region for a given angular separation bin. The uncertainties correspond to the standard error of the mean, i.e. $\sigma_\mu=\sigma/\sqrt{n}$ with $\sigma$ the standard deviation of the population and $n$ the number of mini-regions. 
For the cross-correlation represented by the red circles, the binning was chosen in the same way as in \citet{GON17}, being the circle in  Figure \ref{Fig:xcorr_data} placed at the centre of the corresponding bin.
This is the main data set, which can be directly compared with that used in \citet{GON17} (shown in Figure \ref{Fig:xcorr_data} as grey circles).  
The main difference between the two estimations are the two latest points at large angular scales that are slightly higher in the \citet{GON17} case. The much higher number of independent mini-regions makes us more confident in the current estimation with a more realistic uncertainties. In addition, we estimate again the cross-correlation function using the same procedure described above, but with a half-bin displacement \citep[i.e. by using the same binning size as in][but the bin centre is now shifted of half-bin size with respect to the previous measurement, shown in red]{GON17}.
The cross-correlation measurements are shown as blue circles in Figure \ref{Fig:xcorr_data}.
This additional set of results will provide us the opportunity to study potential dependence of our final results produced by the binning choice (important at large angular scales due to the much lower number of pairs in excess with respect to the random case and the cosmic variance).

\subsubsection{Model}
Similar to the auto-correlation case, and as described in \citet{GON17} and summarised in \citet{BON19}, in order to interpret a foreground-background source cross-correlation signal we adopt the halo model formalism from \citet{COO02}. We define halos as spherical regions whose mean overdensity with respect to the background at any redshift is given by its virial value, which is estimated following \citet{WEI03} assuming a $\Lambda$CDM model with $\Omega_m+\Omega_\Lambda=1$.
The density profile adopted is the one by \citet{NAV96} (hereafter NFW) with the concentration parameter given in \citet{BUL01}.

Weak lensing provides the connection between the low redshift galaxy-mass correlation and the cross-correlation between the foreground and background sources.  The mass density field traced by the foreground galaxy sample causes weak lensing; the weak lensing field is imprinted onto the number counts of the background galaxy sample through magnification bias.
Following mainly \citet{COO02} \citep[see][for details]{GON17}, we compute the correlation between the foreground and background sources adopting the standard Limber \citep{LIM53} and flat-sky approximations \citep[see e.g.][and references therein]{KIL17}. It can be evaluated as
\begin{equation}
    \begin{split}
    w_{fb}=2(\beta -1)\int^{z_s}_0 \frac{dz}{\chi^2(z)}\frac{dN_f}{dz}W^{lens}(z) \\
    \int_{0}^{\infty}\frac{ldl}{2\pi}P_{gal-dm}(l/\chi^2(z),z)J_0(l\theta) 
    \end{split}
,\end{equation}
where
\begin{equation}
    W^{lens}(z)=\frac{3}{2}\frac{H_0^2}{c^2}E^2(z)\int_z^{z_s} dz' \frac{\chi(z)\chi(z'-z)}{\chi(z')}\frac{dN_b}{dz'}
\end{equation}
with $E(z)=\sqrt{\Omega_m(1+z)^3+\Omega_{\Lambda}}$, $dN_b/dz$, and $dN_f/dz$ as the {unit-normalised} background and foreground redshift distribution, and $z_s$ the source redshift. Because $\beta$ is the logarithmic slope of the background sources number counts,   $N(S) = N_0 S^{-\beta}$, and $\chi(z)$ is the comoving distance to redshift z. It is common to assume $\beta=3$ as that leads to the largest gravitational lensing effect in general, and the magnification bias in particular \citep{LAP11, LAP12, CAI13, BIA15, BIA16, GON17, BON19}. In the model $\beta$ is a globally defined constant, but small variations in its value are almost completely compensated by changes in the $M_{min}$ parameter. For example, an increase of 15\% in $\beta$ reduces $\log M_{min}$ in $\sim 1$\%.

The halo model formalism for the cross-correlation function implies that both the low and high redshift galaxy samples trace the same dark matter distribution. The foreground galaxies at the lenses redshift ($z \sim 0.4$) directly trace the dark matter distribution, and the background galaxies ($z >1$) indirectly trace it through the weak lensing effects. Following on from this consideration, the one-halo term relates to the subhalo correlation (by both samples) in the same halo and the two-halo term describes the correlation between one halo by the foreground galaxies and one halo by the background sources.

\subsection{Estimation of parameters}
\label{sec:par_estimate}

To estimate the different sets of parameters, we performed a Markov chain Monte Carlo (MCMC) analysis using the open source {\it emcee} software package \citep{EMCEE}. It is an MIT licensed pure-Python implementation of \citet{GOO10} Affine Invariant MCMC Ensemble sampler. In this analysis we generated at least 10000 posterior samples to ensure a good statistical sampling after convergence.

The estimation of parameters in the foreground auto-correlation function analysis relies on the code\footnote{https://ui.adsabs.harvard.edu/abs/2020ascl.soft02006R/abstract} by \citet{RON20}. We sampled  the space of the model-defining parameters using  MCMC to maximise a likelihood built as the sum of the $\chi^2$ of the two measures we wanted to fit, namely the two-point angular correlation function at a given redshift, $\omega(\theta, z)$, and the average number of sources at a given redshift, $n_g(z)$:
\begin{equation}
    \label{eq:loglike00}
    \log\mathcal{L} \equiv -\dfrac{1}{2}\biggl(\chi^2_{\omega(\theta, z)} + \chi^2_{n_g(z)}\biggr)\ \text{.}
\end{equation}

In the cross-correlation function analysis, we took into account both the astrophysical HOD parameters, and the cosmological parameters. The astrophysical parameters to be estimated are $M_{min}$, $M_1$, and $\alpha$. The cosmological parameters we want to constrain are $\Omega_m$, $\sigma_8$, and $h$. With the current samples, we do not have the statistical power to constrain $\Omega_B$, $\Omega_{\Lambda}$, and $n_s$ in our analysis, and we keep them fixed to \textit{Planck} cosmology, $\Omega_B=0.0486$, $\Omega_{\Lambda}=0.6911$, and $n_s=0.9667$ (see \citealt{PLA18_VI}). A traditional Gaussian likelihood function was used in this case.

It should be noted that only the cross-correlation data in the weak lensing regime ($\theta \ge 0.2$ arcmin) are being taken into account for the fit since we are in the weak lensing approximation \citep[see][for a detailed discussion]{BON19}. Moreover, the foreground galaxy density $n_g$ is not included in the cross-correlation analysis as an additional constraint.

\subsection{Priors of parameters}
\label{sec:par_priors}

The \textit{emcee} package allows us to impose priors on the parameters to be estimated with   MCMC, i.e. it allows   MCMC to explore just the regions delimited by given boundaries. This is usually done in order to avoid spending computing time in searching  physically unacceptable parameter space. In this work we want to study both a prior-free case (by imposing flat priors) and more constrained parameters regions (by applying Gaussian priors).

As for the auto-correlation method performed on the GAMA sample (see section~\ref{sec:auto_corr}), we apply relatively extensive priors to the parameters in order to keep this analysis independent from the one with cross-correlation. In particular, the priors are set to $\log{(M_{min}/M_\odot)}$ and $\log{(M_1/M_\odot)}$ between 9 and 15, $\alpha$ between 0.1 and 1.5, and $A$ between $-5$ and $-1$.

As for the cross-correlation method (see section~\ref{sec:cross_corr}), we set the priors of the astrophysical parameters to min-max values (in the flat priors case) and mean-$\sigma$ values (in the Gaussian priors case)   in agreement with information available in the literature. In particular, after a review of the available literature on the foreground sample characteristics we extract the following flat (FPA) and Gaussian (GPA) priors for our astrophysical parameters.

As described above, the astrophysical parameters involved in our estimations are $M_{min}$, $M_1$, and $\alpha$. In particular, we rely on \citet{SIF15} for $M_{min}$, on both \citet{VIO15} and \citet{SIF15} for $M_1$, and on \citet{VIO15} for $\alpha$.

As for $M_{min}$, our choice of \citet{SIF15} is based on the fact that their sample is similar to ours: it consists of the 100 $deg^2$ overlapping region between KiDS and GAMA surveys, with a redshift range of $0-0.5$ and an average of 0.25. In their work the average galaxy halo mass is determined for a sample of satellite galaxies in massive galaxy groups ($M>10^{13} h^{-1}M_\odot$).
In particular, in their Figure 2 (top right panel) the satellite distributions of stellar mass to the BCG are shown. 
Since   $M_\star$ are the satellite masses, they are a proxy of the minimum mass (galaxies with no satellites), but it has to be expressed in terms of halo mass in order to be considered   a limit on $M_{min}$. We translate these stellar masses in the corresponding mass of the halo using the recipe by \citet{PAN19} to switch from $\log{M_\star (h^{-1}M_\odot)}$ to $\log{M_h}$. \citet{SIF15} gives values of $\log{M_\star (h^{-1}M_\odot)}$ between 9.5 and 11.5, which correspond to the range $\log{M_h}=11.6-13.6$. So, we use these values to set our flat priors on $\log{M_{min}}$.

In order to apply a more constraining range on the allowed values, we also investigate the case with Gaussian priors by relying again on the work by \citet{SIF15}. In particular, we focus on their Table 1; it gives values of $\log{\langle M_{\star} \rangle}(h^{-1}M_\odot)$ between 10.36 -- 10.81, the lowest possible value being in Bin 1 (taking into account the lower error) and the highest possible value in Bin 3 (taking into account the upper error). Such stellar masses translate \citep[using ][]{PAN19} into $\log{M_h} (M_\odot)$ between 12.3 and 12.5. So, we choose a mean value of $\mu=12.4$ and  a dispersion of $\sigma=0.1$ for the Gaussian priors. In this way the $\log{M_{min}}$ priors stay in the \citet{SIF15} more restrictive findings within the 68\% confidence level (CL).

It should be noted that these flat and Gaussian priors are in agreement with \citet{AVE15}. In particular, their Figure 28 shows values of $\log{M_h}(M_\odot)$ between 11.5 -- 13.5, and in their Figure 5 they gave a value of $\log{M_{\star}}(M_\odot)=10.5$ that translates to $\log{M_{h}}(M_\odot)=12.3$.

Concerning $M_1$, we rely on \citet{VIO15} (see their Figure 16) to set the wide flat priors to $\log{M_1}(M_\odot)=13-14.5$. In order to limit this parameter to a more restricted range, we consider again \citet{SIF15} instead. The Gaussian priors are obtained using the information in their Table 2: we choose the range $\log{M_{host}}(h^{-1}M_\odot)=13.51-14.8$, which spans from the lowest possible value (i.e. the one for Bin~1 including the lower error) and the highest possible value (i.e. the one for Bin~3 taking into account the upper error). By writing in the units we are using in this work, we get a typical halo mass with at least one satellite in the range $\log{M_1}(M_\odot)=13.66-14.33$. To set the Gaussian priors based on these values, we take the mean $\mu=13.95$ of this range and a dispersion of $\sigma=0.3$, so that the \citet{SIF15} value corresponds to the 68\%  CL of our priors. We can argue that even smaller Gaussian priors can be set by only relying  on their results for Bin 1 \citep[in][Table 2]{SIF15}. The choice of this bin is justified because  we are looking for the minimum mass to have at least one satellite. However, we decided to use a less constraining prior in order to be more conservative.

With respect to $\alpha$, we derive our priors by comparing our eq. \ref{eq:nsat01} to the expression for $\alpha$ in \citet{VIO15} (eq. 40). From this comparison, we obtain a value of $\alpha=0.92 \pm 0.15$. 
So, we set the Gaussian priors with mean $\mu=0.92$ and dispersion $\sigma=0.15$. To define the flat priors, we consider instead a $3\sigma$ range around the mean value, obtaining $\alpha=0.5-1.37$. We also perform an additional analysis relying on the auto-correlation results, by setting the Gaussian priors to the $\mu$ and $\sigma$ rounded values in Tab.\ref{tab:bestfit}. Finally, we set common flat priors for the Cosmological parameters: 0.1--0.8 for $\Omega_m$, 0.6--1.2 for $\sigma_8$, and 0.5--1.0 for $h$.

\section{Results}
\label{sec:results}

To gather as much information as possible about the derived cosmological constraints and their robustness, first we fix the cosmology to the \textit{Planck} one and study the behaviour of the astrophysical parameters. We decide to use a non-informative, or uniform/flat, priors for our free astrophysical parameters and we perform the analysis for both data sets (the default  and the shifted binning). Then we also allow the cosmological parameters to vary using flat priors for three different astrophysical parameter assumptions: a) flat priors from the literature, b) Gaussian priors from the literature, and c) Gaussian priors from the auto-correlation analysis described in section \ref{sec:auto_corr}.

\begin{table}[ht] 
\caption{Estimated parameter values and uncertainties with the auto-correlation data analysis.} 
\label{tab:bestfit} 
\centering 
\begin{tabular}{c c c c c} 
\hline 
\hline 
Param & priors & \multicolumn{3}{c}{autocorrelation}  \\ 
 & F[a,b] & $\mu$ & $\sigma$ & peak\\ 
 &  & $\pm 68 CL$ & $68CL$ &  \\ 
\hline 
\\   
$\log(M_{min}/M_\odot)$ & F[9,15] & $12.362_{-0.007}^{+0.002}$ & 0.007 & 12.36\\  
\\   
$\log(M_1/M_\odot)$ & F[9,15] & $13.64_{-0.09}^{+0.03}$ & 0.07 & 13.56\\    
\\   
$\alpha$ & F[0.1,1.5] & $1.17_{-0.05}^{+0.06}$ & 0.1 & 1.18\\     
\\   
$ A$ & F[-5,-1] & $-3.08_{-0.05}^{+0.05}$ & 0.07& -3.14\\   
\\   
\hline 
\hline 
\end{tabular} 
\end{table}  

\begin{figure}[ht]
    \centering
    \includegraphics[width=0.45\textwidth]{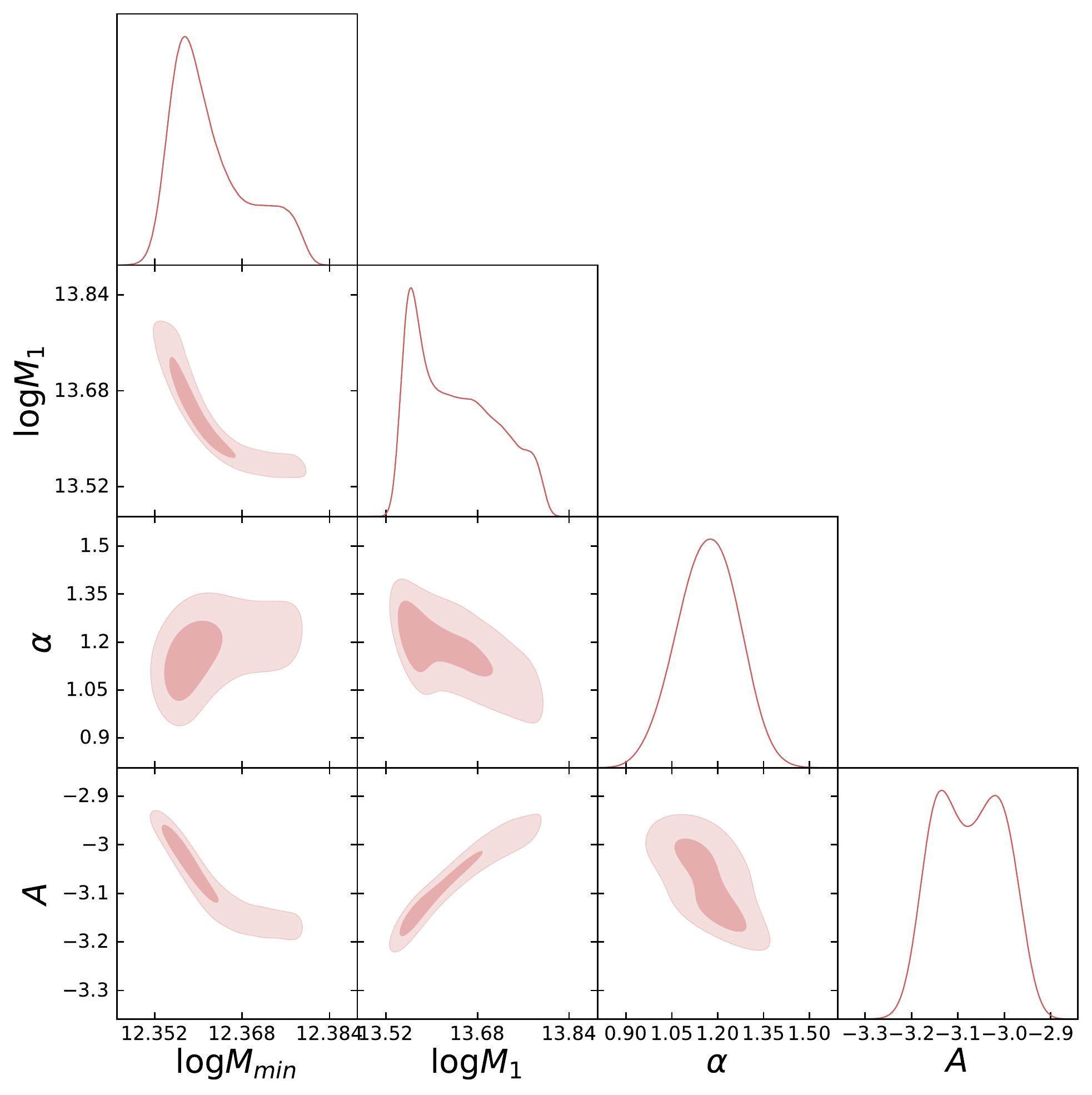}
    \caption{Posterior distributions for the model parameters in the auto-correlation data analysis. The contours for these plots are set to 0.393 and 0.865 (see text for  details).}
    \label{fig:corner}
\end{figure}

\subsection{Astrophysical parameters}
\label{sec:astro_params}

\subsubsection{Auto-correlation data analysis}
\label{sec:astro_params_auto}

\begin{table*}[ht] 
\caption{Results obtained from the cross-correlation data, computed in two different binnings, with flat priors on astrophysical parameters and fixed cosmology. From left to right the columns are the parameters, the priors, and the results for
the default  and the shifted binnings (the mean $\mu$ with   upper and lower limits at   68 \% CL, the $ \sigma$, and the peak of the  posterior distribution). Parameters without a $\sigma$ value indicate that
there is no constraint at 68\% CL, i.e. they are unconstrained.} 
\label{Tab:Run1-2} 
\centering 
\begin{tabular}{c c c c c c c c} 
\hline 
\hline 
Param & Priors & \multicolumn{3}{c}{FPA} & \multicolumn{3}{c}{FPA shifted} \\ 
 & F[a,b] & $\mu$ & $\sigma$ & peak & $\mu$ & $\sigma$ & peak\\ 
 &  & $\pm 68 CL$ & $ 68 CL $ &  & $\pm 68 CL$ & $68 CL$ & \\ 
\hline 
\\   
$\log(M_{min}/M_\odot)$ &  F[11.6,13.6] & $12.42_{- 0.11}^{+ 0.20}$ &  0.20& 12.46 & $12.41_{- 0.07}^{+ 0.17}$ &  0.16 & 12.45\\   
\\   
$\log(M_1/M_\odot)$ & F[13,14.5] & - &  - & 14.12 & $13.65_{- 0.54}^{+ 0.29}$ &  0.38 & 13.59\\   
\\   
$\alpha$ & F[0.5,1.37] & $ 0.89_{- 0.39}^{+ 0.13}$ &  0.24&  0.53 & $ 0.90_{- 0.40}^{+ 0.14}$ &  0.24 &  0.50\\   
\\   
\hline 
\hline 
\end{tabular} 
\end{table*}  

\begin{figure}[ht]
\centering
\includegraphics[width=9cm]{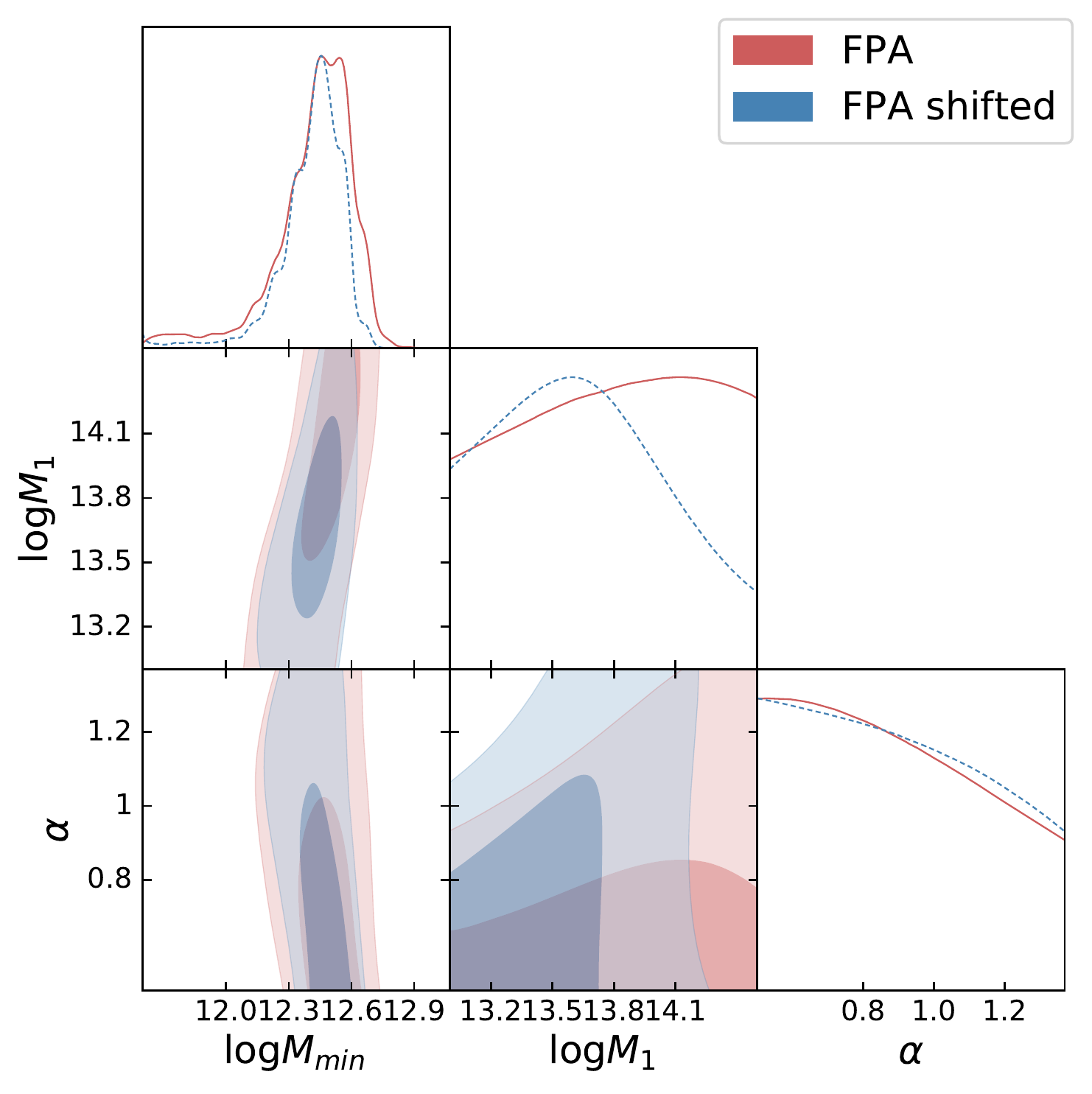}
    \caption{Results obtained from the cross-correlation data with flat priors and fixed cosmology on the cross-correlation data in the two data sets described in Sect. \ref{sec:data} to obtain constraints on the astrophysical parameters. Priors and results are listed in Table \ref{Tab:Run1-2}. The contours for these plots are set to 0.393 and 0.865 (see text for  details).}
    \label{Fig:fpa}
\end{figure}

The posterior distributions for the model parameters (see section \ref{sec:auto_model}) derived from the estimated auto-correlation function (see section \ref{sec:auto_corr}) through the MCMC analysis (see section \ref{sec:par_estimate}) are shown in Fig.~\ref{fig:corner}\footnote{The relevant 1$\sigma$ and 2$\sigma$ levels for a 2D histogram of samples is 39.3\% and 86.5\%, not 68\% and 95\%. Otherwise, there is no  direct comparison with the 1D histograms above the contours.}. 
For this analysis we fix the cosmology to the \textit{Planck} values:   $\Omega_m= 0.3089$, $\sigma_8=0.8159$, and $h=0.6774$.

The mean values at 68\% CL, standard deviations, and  peak values are summarised in Table~\ref{tab:bestfit}. The auto-correlation function calculated using the best-fit parameter values is also shown in Fig.~\ref{fig:acf_model}. The model parameters are well constrained and they provide a good fit to the observed auto-correlation function, obtaining a $\chi^2_\omega$ of 8.836 for 11 degrees of freedom (d.o.f.; corresponding to a reduced $\chi^2_\omega$ of 0.803) and a $\chi^2_{n_g}$ of 0.001 (0.0001). The  value for the $\chi^2_{n_g}$ is so small because we  estimated just one single value and   this   helps in its correct recovery.

Moreover, the values obtained for $\log (M_{min}/M_\odot)$ and $\log (M_1/M_\odot)$ ($12.362^{+0.002}_{-0.007}$ and $13.64^{+0.03}_{-0.09}$, respectively) are consistent with what can be found in  the literature, as discussed in section \ref{sec:astro_params} ($\log (M_{min}/M_\odot)=11.6-13.6$ and $\log (M_1/M_\odot)=13.0-14.5 $ for the flat priors cases). The derived $\alpha$ parameter ($1.17^{+0.06}_{-0.05}$) is also in agreement with the literature values ($0.5-1.37 $ for the flat priors case), although slightly toward the upper limit.

Therefore, we can conclude that the auto-correlation function analysis results can be an additional alternative set of Gaussian parameters for the astrophysical parameters to be used in the full cosmological analysis.

\subsubsection{Cross-correlation data analysis}

\begin{table*}[ht] 
\caption{Results obtained from the cross-correlation data with flat priors (FPA FPC; second column) on astrophysical and cosmological parameters. From left to right, the columns are the parameters, the priors, and the results (the mean $\mu$ with the upper and lower limit at the 68 \% CL, the $ \sigma$, and the peak of the  posterior distribution) for the default  and   shifted binning, respectively. Parameters without a $\sigma$ value indicate that there is no constraint at 68\% CL, i.e. they are unconstrained.} 
\label{Tab:Run3-4} 
\centering 
\begin{tabular}{c c c c c c c c} 
\hline 
\hline 
Param & Priors & \multicolumn{3}{c}{FPA FPC} & \multicolumn{3}{c}{FPA FPC shifted} \\ 
 & F[a,b] & $\mu$ & $\sigma$ & peak & $\mu$ & $\sigma$ & peak\\ 
 &  & $\pm 68 CL$ & $ 68 CL $ &  & $\pm 68 CL$ & $68 CL$ & \\ 
\hline 
\\   
$\log(M_{min}/M_\odot)$ & F[11.6,13.6] & $12.53_{- 0.16}^{+ 0.29}$ &  0.27& 12.61 & $12.50_{- 0.17}^{+ 0.23}$ &  0.22 & 12.53\\   
\\   
$\log(M_1/M_\odot)$ & F[13,14.5] & - &  - & 14.39 & $13.74_{- 0.51}^{+ 0.42}$ &  0.40 & 13.69\\   
\\   
$\alpha$ & F[0.5,1.37] & - &  - &  0.50 & $ 0.89_{- 0.39}^{+ 0.13}$ &  0.24 &  0.50\\   
\\   
$\Omega_m$ & F[0.1,0.8] & $ 0.54_{- 0.08}^{+ 0.26}$ &  0.16&  0.67 & $ 0.52_{- 0.09}^{+ 0.28}$ &  0.17 &  0.59\\   
\\   
$\sigma_8$ & F[0.6,1.2] & $ 0.78_{- 0.15}^{+ 0.07}$ &  0.11&  0.74 & $ 0.80_{- 0.17}^{+ 0.09}$ &  0.12 &  0.77\\   
\\   
$h$ & F[0.5,1.0] & $ 0.76_{- 0.08}^{+ 0.24}$ &  0.14&  1.00 & - & - &  0.85\\   
\\   
\hline 
\hline 
\end{tabular} 
\end{table*}  

\begin{figure*}[ht]
\centering
\includegraphics[width=\textwidth]{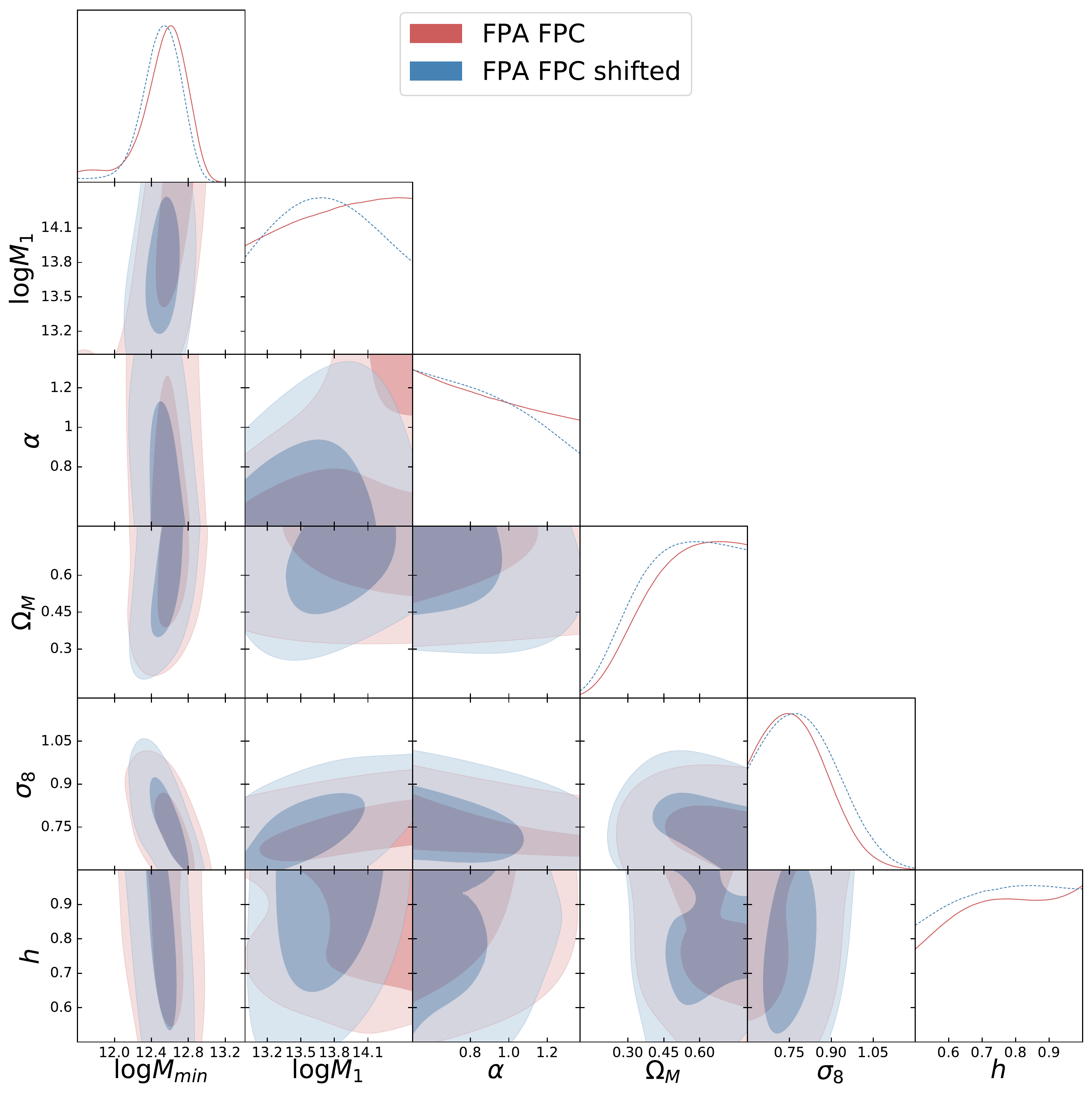}
    \caption{Results obtained from the cross-correlation data with flat priors on astrophysical and cosmological parameters, in the two data set for the cross-correlation data. The limits on the priors for the astrophysical and cosmological parameters and the corresponding results are listed in Table \ref{Tab:Run3-4}. The contours for these plots are set to 0.393 and 0.865 (see text for  details).}
    \label{Fig:fpa_fpc}
\end{figure*}

We perform a similar analysis as in the auto-correlation case for the cross-correlation function. We also fix the cosmology to the \textit{Planck} values, and perform the MCMC analysis on the astrophysical parameters only.
 
The mean values, CL, standard deviations, and the peak values are summarised in Table~\ref{Tab:Run1-2}. The results of the MCMC runs for the two shifted data sets are listed in the last two columns. The posterior distributions shown in Figure~\ref{Fig:fpa} give compatible results to the auto-correlation case (summarised in Table~\ref{tab:bestfit}), in particular for $M_{min}$, the only one that is well constrained.

The value of $M_{min}$ is well constrained with both data sets (default and shifted binning) with almost identical results ($\log (M_{min}/M_\odot)=12.42_{- 0.11}^{+ 0.20}$ and $12.41_{- 0.07}^{+ 0.17}$, respectively) and in agreement with the results from \citet{GON17} and \citet{BON19}, and to typical values in the literature, as described in section \ref{sec:astro_params}.

It is  worth mentioning here  that matter distribution at z<0.2 is not significantly biasing our lens sample (because of lensing, see section~\ref{sec:fore_sample}). This is confirmed because the mass we estimate with the cross-correlation for our selected GAMA sample is in fairly good agreement with that found by the auto-correlation in the GAMA sample (section~\ref{sec:astro_params_auto}), where only very few sources might be affected by lensing and the mass determination is dominated by the unlensed sources. Since the cross- and auto-correlation results are in agreement, it means that  cross-correlation results are also not affected by magnification due to z<0.2 matter distribution.

On the contrary, the  $M_1$ and $\alpha$ parameters are both unconstrained with the two data sets. The results are mostly the mean values between the flat prior limits as expected in this situation. However, there is a hint in the shifted data set toward a similar $\log (M_1/M_\odot$) value  (with a peak value of 13.64), but lower $\alpha$ value ($\sim 0.9$ ) with respect to the auto-correlation case.

Figure \ref{Fig:sensitivity} is of great help in understanding this question. Once the $M_{min}$ is fixed, for the flat ranges adopted for both $M_1$ and $\alpha$, the variation is small enough to avoid a clear constraint, even more if we consider that  both parameters are probably degenerate to a certain degree. 
Moreover, focusing on $M_1$, the analysis based on the shifted cross-correlation function is more constraining than the other one (see e.g. Figure~\ref{Fig:fpa_fpc}). What is causing the difference between these two cases might probably be the first cross-correlation point at $\theta \gtrsim 0.2$ arcmin which appears to be lower in the shifted case than in the other. Looking again at Figure~\ref{Fig:sensitivity}, it seems that lower values of $M_1$ predict lower amplitude of the cross-correlation function at those scales.

For these reasons, we decided to consider these results simply as a self-consistency check of the astrophysical parameters outputs. For the full cosmological analysis, we restricted ourselves to the priors discussed before and the auto-correlation results.

Finally, taking into account the results, we can also conclude that they are robust against small variations in the data set used.

\subsection{Cosmological parameters}
\label{sec:cosmo_params}

\begin{table*}[ht] 
\caption{Results obtained from the cross-correlation data with Gaussian astrophysical and flat cosmological priors (GPA FPC; second column). The column information is the same as in Table \ref{Tab:Run3-4}. In the second column, we list the priors as G($\mu$, $\sigma$) and F[a,b] for the Gaussian and flat priors, respectively.}
\label{Tab:Run5-6} 
\centering 
\begin{tabular}{c c c c c c c c} 
\hline 
\hline 
Param & Priors & \multicolumn{3}{c}{GPA FPC} & \multicolumn{3}{c}{GPA FPC shifted} \\ 
 & G($\mu$,$\sigma$) & $\mu$ & $\sigma$ & peak & $\mu$ & $\sigma$ & peak\\ 
 & F[a,b] & $\pm 68 CL$ & $ 68 CL $ &  & $\pm 68 CL$ & $68 CL$ & \\ 
\hline 
\\   
$\log(M_{min}/M_\odot)$ & G(12.4,0.1) & $12.44_{- 0.09}^{+ 0.10}$ &  0.09& 12.43 & $12.43_{- 0.09}^{+ 0.09}$ &  0.09 & 12.43\\   
\\   
$\log(M_1/M_\odot)$ & G(13.95,0.30) & $13.92_{- 0.25}^{+ 0.29}$ &  0.26& 13.95 & $13.89_{- 0.27}^{+ 0.28}$ &  0.27 & 13.89\\   
\\   
$\alpha$ & G(0.92,0.15) & $ 0.92_{- 0.15}^{+ 0.15}$ &  0.15&  0.92 & $ 0.90_{- 0.15}^{+ 0.15}$ &  0.15 &  0.90\\   
\\   
$\Omega_m$ & F[0.1,0.8] & $ 0.51_{- 0.16}^{+ 0.22}$ &  0.17&  0.57 & $ 0.52_{- 0.13}^{+ 0.23}$ &  0.16 &  0.51\\   
\\   
$\sigma_8$ & F[0.6,1.2] & $ 0.84_{- 0.10}^{+ 0.10}$ &  0.10&  0.86 & $ 0.87_{- 0.10}^{+ 0.10}$ &  0.10 &  0.87\\   
\\   
$h$ & F[0.5,1.0] & $ 0.79_{- 0.07}^{+ 0.21}$ &  0.13&  1.00 & $ 0.78_{- 0.07}^{+ 0.22}$ &  0.14 &  1.00\\   
\\   
\hline 
\hline 
\end{tabular} 
\end{table*}

\begin{figure*}[ht]
\centering
\includegraphics[width=\textwidth]{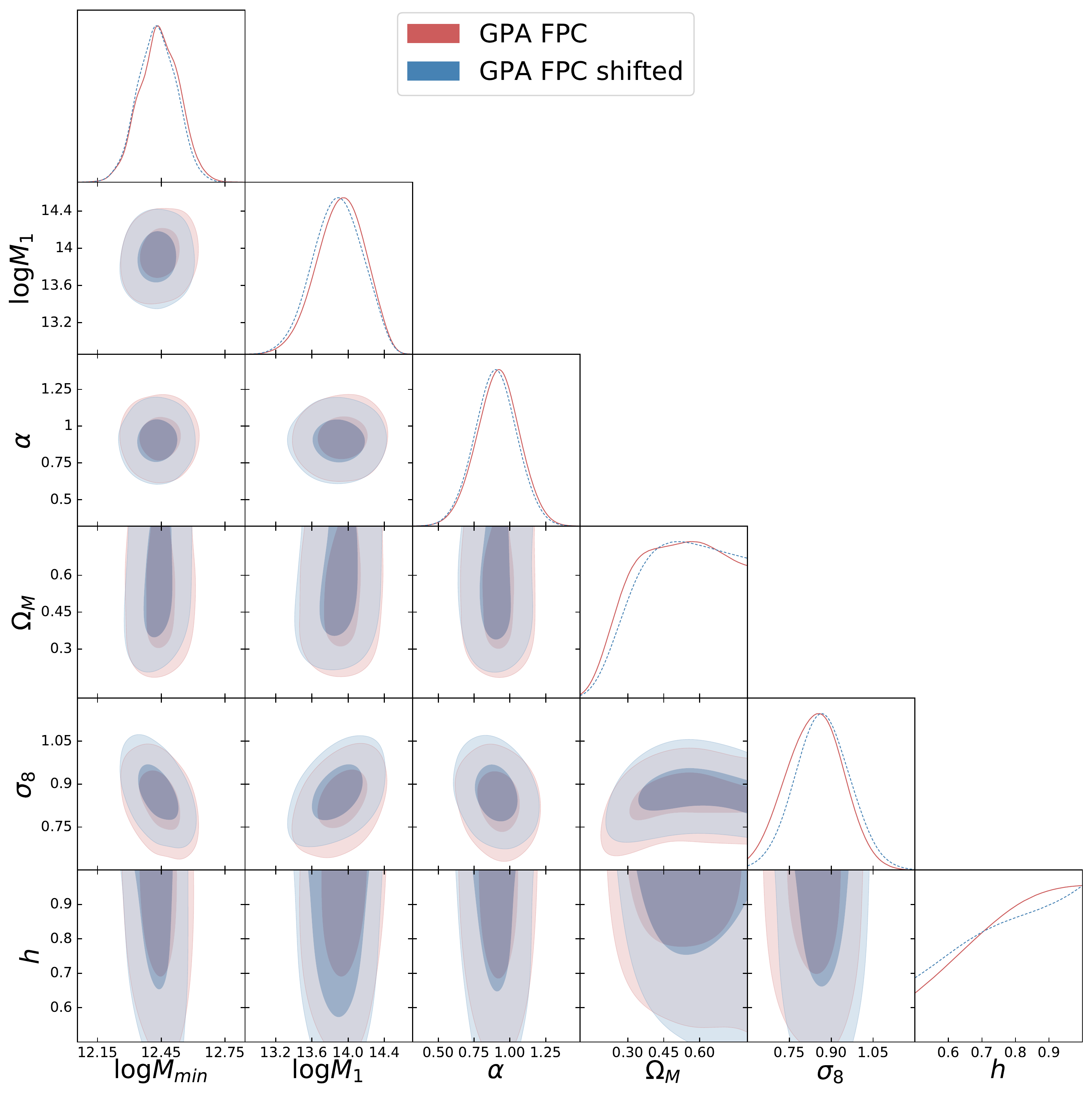}
    \caption{Results obtained from the cross-correlation data with Gaussian astrophysical and flat cosmological priors in the two shifted data sets for the cross-correlation data. The results are listed in Table \ref{Tab:Run5-6}. The Gaussian priors are set to the $\mu$ and $\sigma$ values presented by \citet{VIO15}, \citet{SIF15}, and \citet{PAN19}. The contours for these plots are set to 0.393 and 0.865 (see text for  details).}
    \label{Fig:gpa_fpc}
\end{figure*}

Having shown that using flat priors on astrophysical parameters provides values compatible with the literature, we extend the cross-correlation function analysis to allow for variations in the main cosmological parameters, $\Omega_m$, $\sigma_8$, and $H_0$. We would like to note at this point that the main aim of this work is to exploit the cross-correlation measurements in order to constrain these parameters using flat priors on both astrophysical and cosmological parameters (FPA FPC). So we  first present our results for this more general case, and then investigate the effects and possible improvements on estimating our cosmological parameters  when setting the astrophysical parameters to the Gaussian priors derived from the literature and from auto-correlation results.

The best-fit posterior distributions of the free parameters is shown in Fig. \ref{Fig:fpa_fpc} (red for the default binning and blue for the shifted). The  best-fit values are listed in Table \ref{Tab:Run3-4}.  
The solid lines in Figure \ref{Fig:xcorr_data} represent the halo model best fit, estimated using the MCMC approach. In particular, the red is for the main data set, which gives a $\chi^{2}$ of 8.16 for 6 d.o.f (corresponding to a reduced $\chi^{2}$ of 1.36; the d.o.f. are the 12 data points minus the 6 parameters) and the blue is for the shifted data set, with a $\chi^{2}$ of 9.92 (1.65, when taking into account the d.o.f.). The green line corresponds to the parameters obtained running on the main data set except for $\Omega_m$ that is set equal to 0.3. In this case, we get $\chi^{2}$ of 13.76 (2.29).

First, the recovered astrophysical parameters are compatible with the results of the previous section (see Table \ref{Tab:Run1-2}). This also means that only $M_{min}$ is well constrained.
Therefore, the conclusion that the results are robust against small variations in the data sets can also be extended   to the cosmological case.

With respect to the cosmological parameter results, they are not well constrained in this proof of concept, but there are interesting conclusions that can be drawn. The $H_0$ parameter is unconstrained, but for $\sigma_8$ there is a clear broad peak around 0.75 and its posterior distribution indicates an upper limit of <1.0 at the 95\% CL.
The analysis of $\Omega_m$ is more complicated, but interesting at the same time. As clearly shown by Fig. \ref{Fig:sensitivity}, $\Omega_m$ is most sensitive to the cross-correlation measurements on the largest scales, where the uncertainties are largest (the three last observations are almost upper limits). This translates into a wide posterior distribution toward higher $\Omega_m$ values driven mostly by the lower value of the last point. On the contrary, there is also a strong constraint on the lower $\Omega_m$ value. This analysis implies that $\Omega_m \gtrsim 0.24$ at the 95\% CL.

In order to investigate possible improvements on the cosmological constraints, we perform additional analyses with Gaussian priors on the astrophysical parameters and flat priors on the cosmological parameters. 

The first choice for the Gaussian priors is the values from the literature described in Sect. \ref{sec:astro_params}. The priors for these runs are listed in the second column of Table \ref{Tab:Run5-6}. The remaining columns list the results of the MCMC for both data sets (same as Table \ref{Tab:Run3-4}). The best-fit posterior distributions of the free parameters are shown in Fig. \ref{Fig:gpa_fpc} (red for the default binning and blue for the shifted).

\begin{figure*}[ht]
\centering
\includegraphics[width=\textwidth]{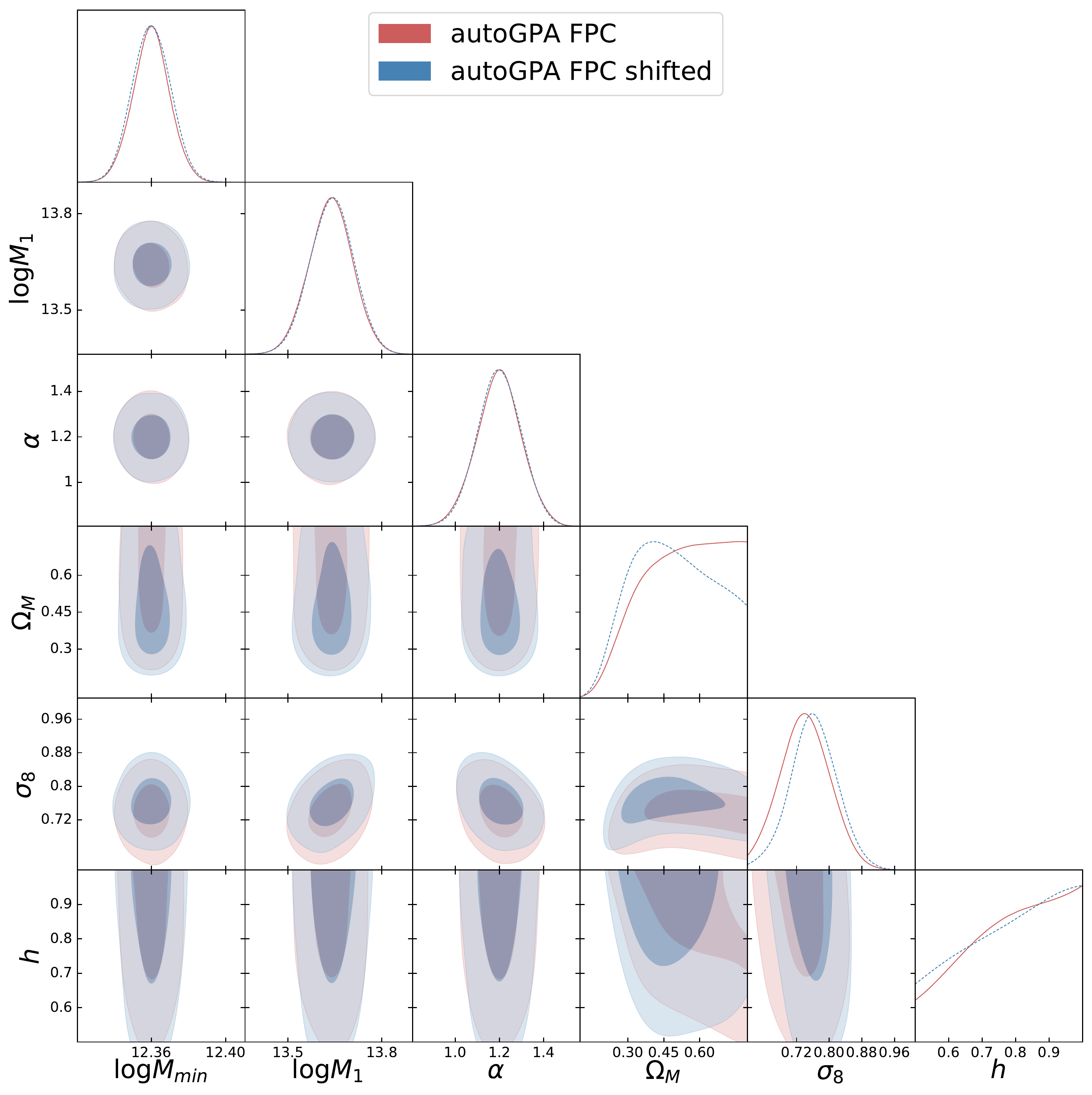}
    \caption{Results obtained from the cross-correlation data with Gaussian astrophysical priors from auto-correlation analysis and flat cosmological priors in the two shifted data sets for the cross-correlation data. The results are listed in Table \ref{Tab:Run7-8}. The contours for these plots are set to 0.393 and 0.865 (see text for  details).}
    \label{Fig:autogpa_fpc}
\end{figure*}

We also analyse the case with Gaussian priors on the astrophysical parameters fixed with the auto-correlation analysis results described in section \ref{sec:auto_corr}. Our findings in this case are summarised in Table \ref{Tab:Run7-8} and Figure \ref{Fig:autogpa_fpc}. 

As expected, the astrophysical parameters are prior dominated. In addition, by imposing Gaussian priors to the astrophysical parameters the uncertainties in the cosmological parameters are reduced. This is clearly shown by comparing the results using the more restrictive priors from the auto-correlation analysis with the Gaussian priors from the literature. For example, the $\sigma_8$ parameter improves from a $\sigma_8=0.84_{- 0.10}^{+ 0.10}$ (Gaussian prior from the literature) to $\sigma_8=0.74_{-0.03}^{+ 0.03}$ (Gaussian prior from the auto-correlation analysis). In both cases there is also a slight correlation between $\sigma_8$ and $M_1$ parameters as both affect the cross-correlation function in an opposite way on intermediate angular scales (see Figure \ref{Fig:sensitivity}).

Both $\Omega_m$ and $H_0$ remain almost unconstrained, as in the flat prior case. There is slight tendency of $H_0$ toward values above $70\, km/s\, Mpc^{-1}$, as is  discussed in more detail in the next section. The strong lower limit constraint found with the flat priors for $\Omega_m$ is confirmed by  additional analysis. Moreover, using the auto-correlation Gaussian priors, the posterior distribution of $\Omega_m$ shows a peak around 0.41 in the shifted case. This is an indication that reducing the uncertainties in the observed cross-correlation function and accurate astrophysical priors can provide strong independent constraints based on the magnification bias.
    
\begin{table*}[ht] 
\caption{Results obtained from the cross-correlation data with Gaussian (from the auto-correlation analysis) astrophysical and flat cosmological priors (autoGPA FPC; second column). The column information is the same as in Table \ref{Tab:Run5-6}}.
\label{Tab:Run7-8} 
\centering 
\begin{tabular}{c c c c c c c c} 
\hline 
\hline 
Param & Priors & \multicolumn{3}{c}{autoGPA FPC} & \multicolumn{3}{c}{autoGPA FPC shifted} \\ 
 & G($\mu$,$\sigma$) & $\mu$ & $\sigma$ & peak & $\mu$ & $\sigma$ & peak\\ 
 & F[a,b] & $\pm 68 CL$ & $ 68 CL $ &  & $\pm 68 CL$ & $68 CL$ & \\ 
\hline 
\\   
$\log(M_{min}/M_\odot)$ & G(12.36,0.01)  & $12.36_{- 0.01}^{+ 0.00}$ &  0.01& 12.36 & $12.36_{- 0.01}^{+ 0.00}$ &  0.01 & 12.36\\ 
\\   
$\log(M_1/M_\odot)$ & G(13.64,0.07) & $13.64_{- 0.04}^{+ 0.04}$ &  0.07& 13.64 & $13.64_{- 0.04}^{+ 0.04}$ &  0.07 & 13.64\\
\\ 
$\alpha$ & G(1.2,0.1) & $ 1.20_{- 0.05}^{+ 0.05}$ &  0.10&  1.20 & $ 1.20_{- 0.05}^{+ 0.05}$ &  0.10 &  1.19\\  
\\ 
$\Omega_m$ & F[0.1,0.8] & $ 0.53_{- 0.06}^{+ 0.27}$ &  0.16&  0.78 & $ 0.49_{- 0.16}^{+ 0.03}$ &  0.16 &  0.41\\  
\\
$\sigma_8$ & F[0.6,1.2] & $ 0.74_{- 0.03}^{+ 0.03}$ &  0.06&  0.74 & $ 0.76_{- 0.03}^{+ 0.03}$ &  0.06 &  0.76\\
\\
$h$ & F[0.5,1.0] & $ 0.79_{-0.06}^{+ 0.21}$ &  0.13&  1.00 & $ 0.79_{-0.06}^{+ 0.21}$ &  0.14 &  1.00\\ 
\\   
\hline 
\hline 
\end{tabular} 
\end{table*} 

\subsection{Comparison with other results}

\begin{figure}[ht]
\centering
\includegraphics[width=8cm]{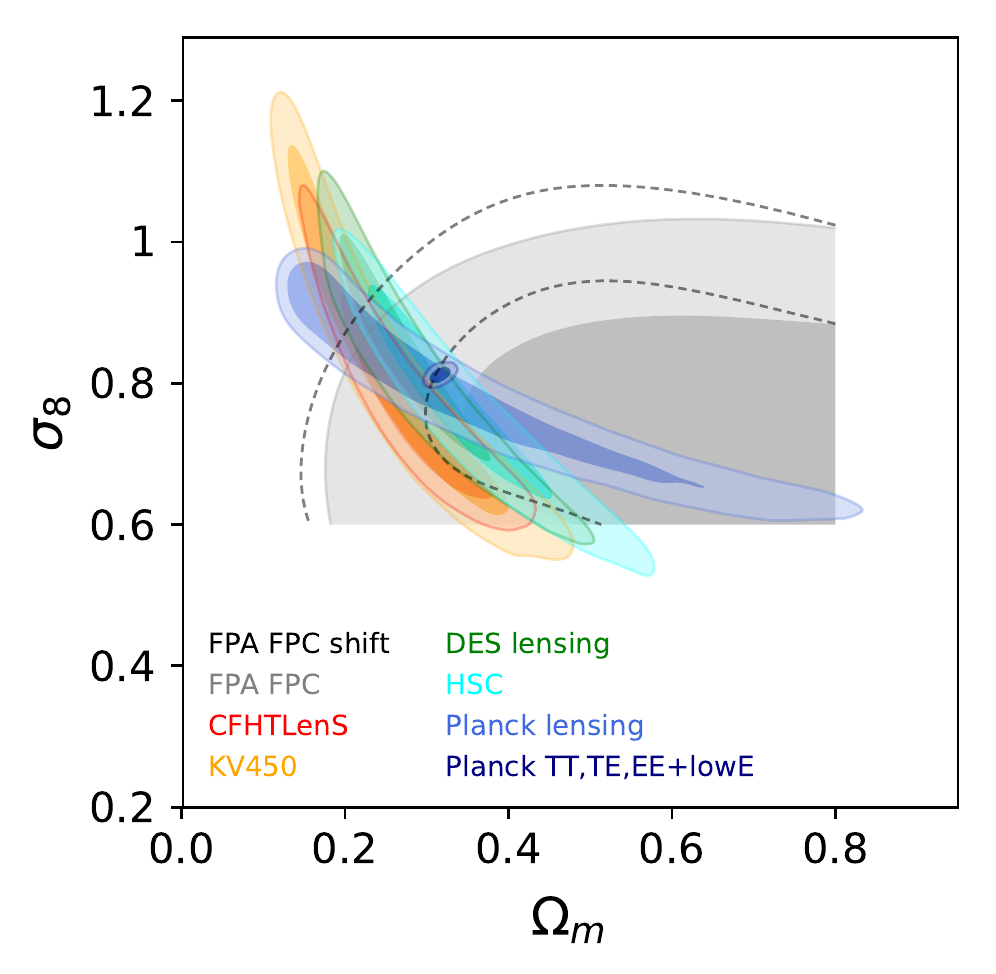}
\caption{$\Omega_{m}$ - $\sigma_{8}$ plot comparing our results using flat priors in  the astrophysical and cosmological parameters (both cases, grey and black) with results from \textit{Planck} (blue and dark blue for the lensing only and all \textit{Planck} cases, respectively), CFHTLenS (red), KV450 (orange), DES lensing (green), and HSC (cyan).
The contours for these plots are set to 0.68 and 0.95.}
    \label{Fig:omegam_sigma8}
\end{figure}

\begin{figure*}[ht]
\centering
\includegraphics[width=8cm]{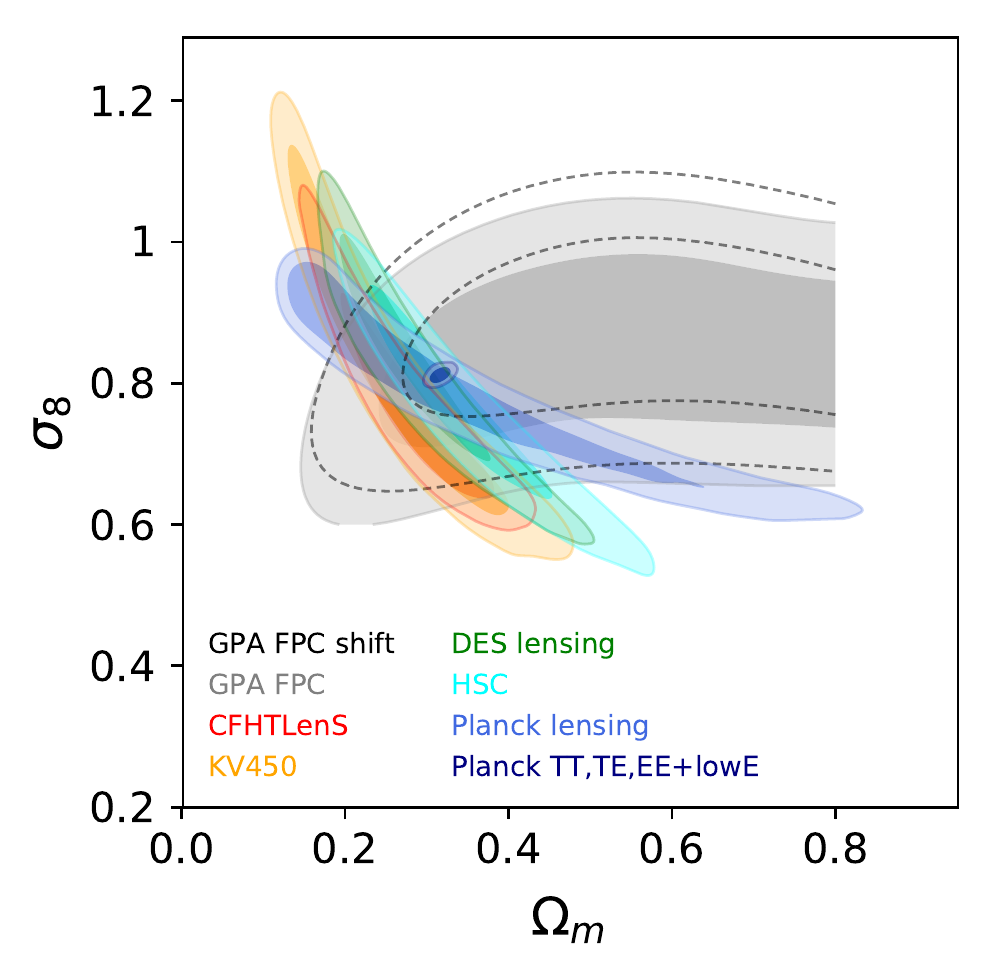}
\includegraphics[width=8cm]{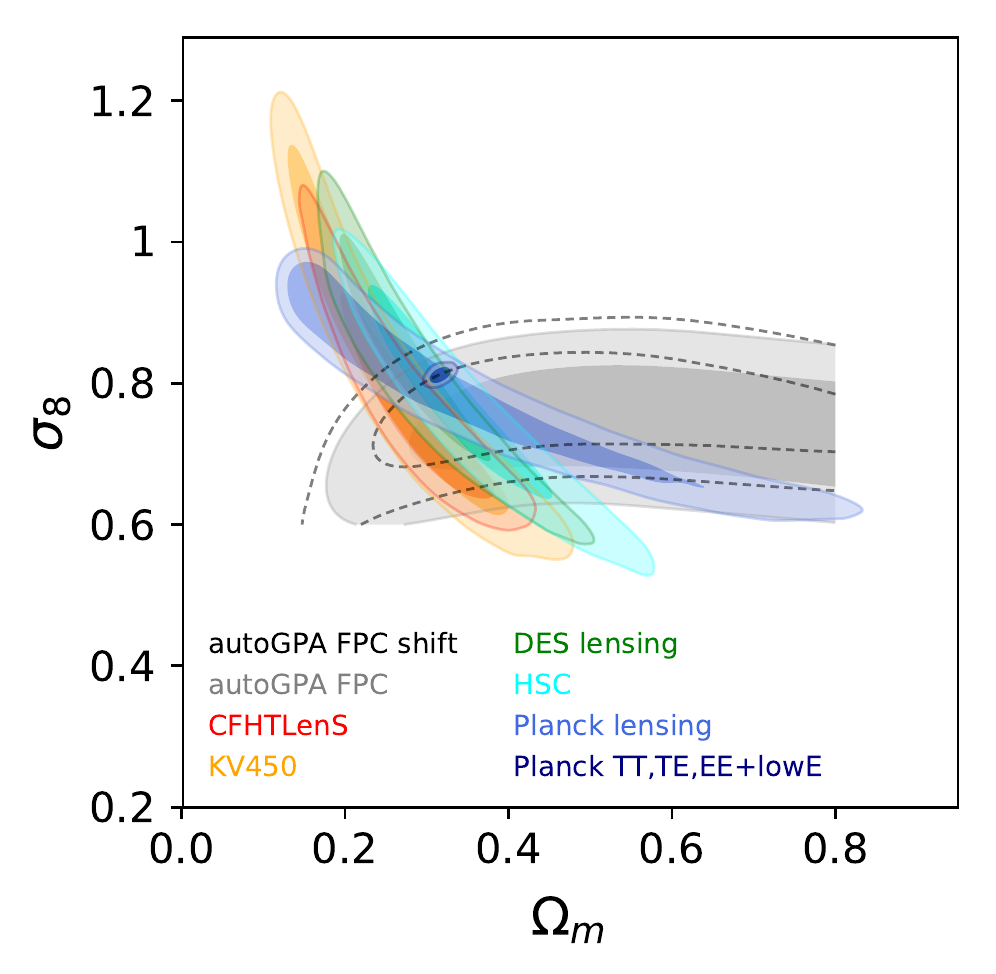}
\caption{$\Omega_{m}$ - $\sigma_{8}$ plot comparing our results using Gaussian priors (left panel: from the literature; right panel: from the auto-correlation analysis) in the cosmological parameters (both cases, grey and black) with results from \textit{Planck} (blue and dark blue for the lensing only and all \textit{Planck} cases, respectively), CFHTLenS (red), KV450 (orange), DES lensing (green), and HSC (cyan). The contours for these plots are set to 0.68 and 0.95.}
    \label{Fig:omegam_sigma8_gpa}
\end{figure*}

\begin{figure}[ht]
\centering
\includegraphics[width=8cm]{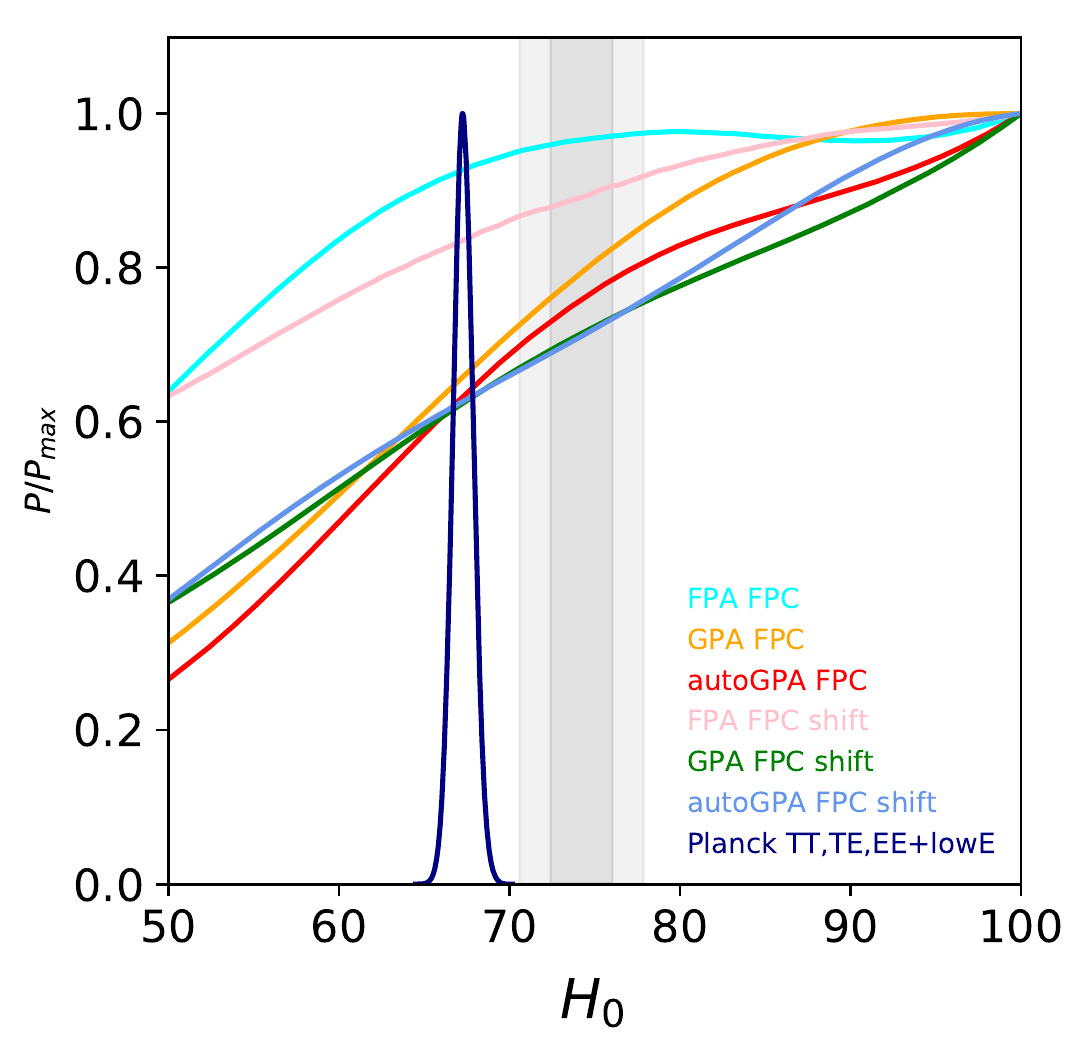}\\
\caption{Posterior distributions on $H_0$ derived using the different data sets and assuming different priors for the astrophysical parameters. As a comparison the result by \citet{RIE19} (grey band) and \textit{Planck} are also shown.}
    \label{Fig:H0_1D}
\end{figure}

In this section we compare our findings with the cosmological constraints obtained using a different observable: the shear of the weak gravitational lensing. We consider both the CMB lensing from \textit{Planck} \citep{PLA18_VIII} and the cosmic shear of galaxies as measured by several surveys, focusing on those that are the most constraining to date. Specifically, we include results from the revised analysis of the cosmic shear tomography measurements of the Canada-France-Hawaii Telescope Lensing Survey presented in CFHTLenS \citep{JOU17}, the first combined cosmological measurements of the Kilo Degree Survey and VIKING based on $450\,$deg$^2$ data \citep[KV450,][]{HIL20}, the first-year lensing data from the Dark Energy Survey (DES, \citet{TRO18}), together with recent constraints from the two-point correlation functions of the Subaru Hyper Suprime-Cam first-year data \citep[HSC,][]{HAM20}. 
It is worth noting that these results span somewhat different redshift ranges and are affected by completely different systematic effects with respect to the magnification bias measurements analysed in this paper, which thus represent a complementary probe to weak lensing.

We focus on the comparison to the constraints in the $\Omega_m$ - $\sigma_8$ plane since weak lensing is most sensitive to these cosmological parameters. In particular, cosmic shear approximately constrains the combination $\sigma_8 \Omega_m^{0.5}$, while $\sigma_8 \Omega_m^{0.25}$ is what is measured by CMB lensing. These combinations identify degeneracy directions that are clearly visible in Figures \ref{Fig:omegam_sigma8} and \ref{Fig:omegam_sigma8_gpa}, where we show the marginalised posterior contours ($68\%$ and $95\%\,$CL)  for the above-mentioned data sets. We use MCMC results that have been publicly released by the different collaborations. It should be noted that in these comparative figures, in order to be consistent with the results shown in the literature for the experiments mentioned above, we set the contours to 0.68 and 0.95, which  are different from the values of 0.393 and 0.865   used in the corner plots shown in the previous sections, and that actually corresponds to the relevant 1$\sigma$ and 2$\sigma$ levels for a 2D histogram of samples. The choice of 0.68 and 0.95 contours in this subsection is  for a direct comparison with the literature; the 0.393 and 0.865 selection in the 2D histograms of the corner plots is for a direct comparison with the 1$\sigma$ and 2$\sigma$ levels in the 1D histograms in the upper part of the same corner plots.

It should be noted that different studies adopt different cosmological parameters priors; however, we do not attempt to adjust them to our fiducial set-up as an in-depth comparison is beyond the scope of this paper. For completeness, in the figures we also show results from \textit{Planck} CMB temperature and polarisation angular power spectra (dark blue). They have been known to be in tension with the CFHTLenS (red) and KV450 (orange) data, while a better agreement has instead been found with the displayed HSC (cyan) and DES (green) constraints. For the last, in particular, \citet{TRO18} made the conservative choice of excluding the non-linear scales where modelling uncertainties are higher. This  also resulted in a reduction of the constraining power of the data set. 

The cosmological constraints derived in this paper are shown for both of the data sets (grey filled contours and black dashed curves). The main results, based on the flat prior for both the astrophysical and cosmological parameters, are shown in Figure \ref{Fig:omegam_sigma8}. In Figure \ref{Fig:omegam_sigma8_gpa}, we show how the results can be improved using Gaussian priors on the astrophysical parameters.
Even if our data sets might not seem constraining, two important conclusions can be drawn from the results. Firstly, the estimates provide a lower bound $\Omega_m > 0.24$ at $95\%$ CL that is robust against the choice of priors. Secondly, our results do not show the typical degeneracy that characterises the cosmic shear results.
This means that future more constraining magnification bias samples can offer a complementary probe able to break the degeneracy (see e.g. Figure \ref{Fig:omegam_sigma8}). 

The results on $\Omega_m$ can be better understood looking at Figure \ref{Fig:sensitivity} in the Appendix, which shows how the cross-correlation function depends on cosmological parameters. It is interesting to see how the largest angular separations are very sensitive to the matter density parameter, and how present data disfavour low values of $\Omega_m$.

We note that the data in this paper alone cannot place useful constraints on the Hubble constant, as  is evident from Figure \ref{Fig:H0_1D}. This is usually the case even for weak lensing measurements. However, in most of our cases there is a mild pull towards higher $H_0$ values. 

As a comparison, we also add the vertical grey band, corresponding to the \citet{RIE19} results and the \textit{Planck} constraints. However, due to our large error bars on the cross-correlation data, our results are not competitive with those in the recent literature and cannot give any hint on the \textit{Planck}--\citet{RIE19} tension problem yet.

\subsection{Future developments}

This work is meant to show that the magnification bias can become an additional independent cosmological probe to help to resolve the current tensions in modern Cosmology. Results are not yet competitive with current estimation of cosmological parameters, but some improvements are expected if smaller error-bars on the cross-correlation measurements can be achieved, especially on larger scales where the $\Omega_m$ and $H_0$ parameters  mostly affect the cross-correlation function. 

First of all, it should be noted that this work is based on just one (relatively large) redshift bin.
As shown in \citet{GON17}, a tomographic analysis can be done with such cross-correlation measurements. When we divide the sample into redshift bins, we  reduce the the sample size and we expect the error bars to increase. This clear disadvantage might be counterbalanced by easier constraining of the cosmological parameters, since only sources with very similar redshift and the known evolution of the cosmological parameters are being taken into consideration. This is a possibility that needs to be further explored.

Regarding the reduction of the error bars, the statistics can be increased enlarging both the lenses and the background source samples. As for the lenses, the sample might be increased with current data by considering sources with photometric redshifts. On the one hand, the number of sources with photometric redshifts is larger than that of sources with the spectroscopic redshifts. On the other hand, photometric redshifts are less accurate, and in this case the application of the cross-correlation that needs very clearly redshift-separated samples of foreground and background sources   might not be straightforward. In the future more precise catalogues of lens sources might be available, such as from DES \citep{DES16}, the Javalambre Physics of the Accelerating Universe Astrophysical Survey \citep[J-PAS,][]{BEN14} and the Euclid \citep{LAU11} experiments/missions.

 For  the background sources, the number of SMGs might be increased by considering the catalogue of the whole area covered by Herschel \citep[e.g. by investigating the possibility of using the Herschel Extragalactic Legacy Project, HELP;][]{SHI19}, and not just H-ATLAS data. Moreover, it will  surely be increased by catalogues of new instruments in the near-infrared, such as the James Webb Space Telescope \citep[JWST;][]{GAR06}.

\section{Conclusions}
\label{sec:conclusion}

The aim of this work was to test the capability of magnification bias with high-z SMGs as a cosmological probe. As demonstrated in \citet{GON17} and \citet{BON19}, magnification bias is a powerful tool for constraining the free parameters of the HOD model. We now exploit cross-correlation data to constrain not only astrophysical parameters ($M_{min}$, $M_1$, and $\alpha$), but also some of the cosmological ones. In particular, we realised that at this early stage the $n_s$, $\Omega_B$, and $\Omega_{\Lambda}$ parameters cannot be constrained. We thus focus our analysis on $\Omega_m$, $\sigma_8$, and $H_0$ for this proof of concept.

With our current results $\Omega_m$ and $H_0$ cannot be well constrained, although we showed that magnification bias can be a way to set constraints on these parameters, but we conclude that the most important issue regarding the $\Omega_m$ and $H_0$ parameters is the uncertainty on the cross-correlation data on large scales. 
So our future efforts must be focused on reducing the error bars on the data set. What we need is a larger area for the background sources to get a wider area and achieve a better statistics.

Moreover, we do not find a degeneracy between $\Omega_m$ and $\sigma_8$. If any, the correlation might be opposite to the common one (see Figure \ref{Fig:omegam_sigma8_gpa}): lower $\sigma_8$ implies in our case lower $\Omega_m$ in order to compensate the effect on large scales.

With all the cases studied in this work we can set a lower limit of 0.24 at the 95\% CL on $\Omega_m$, and we see a mild preference towards high $H_0$ values.
Since $\Omega_m$ was allowed to vary in the range [0.1-0.8], this lower bound is a very robust conclusion and its importance is clearly shown in the comparison with the results from weak lensing data: our lower bound still imposes restrictions to the other weak lensing results. 

For our constraints on $\sigma_8$, we obtained only a tentative peak around 0.75 but an interesting upper limit of $\sigma_8\lesssim 1$ at 95\% CL. By imposing Gaussian priors to the astrophysical parameters, the constraints are improved: $\sigma_8=0.74_{- 0.03}^{+ 0.03}$ using the Gaussian prior from the auto-correlation analysis.

Finally, our estimate of the $H_0$ parameter is still not accurate, being inclusive of the \textit{Planck} and the \citet{RIE19} results. Better results might be obtained if better measurements and smaller error bars for the cross-correlation points on large scales can be achieved.

\begin{acknowledgements}
The authors greatly thanks the anonymous referee for providing precious and appropriated comments.
LB, JGN, MMC acknowledge the PGC 2018 project PGC2018-101948-B-I00 (MICINN/FEDER). LB and JGN acknowledge PAPI-19-EMERG-11 (Universidad de Oviedo). 
MM is supported by the program for young researchers ``Rita Levi Montalcini" year 2015. 
LD and SJM acknowledge support from the European Research Council Advanced Investigator grant, COSMICISM and Consolidator grant, COSMIC DUST. 
AL acknowledges partial support from PRIN MIUR 2015 Cosmology and Fundamental Physics: illuminating the Dark Universe with Euclid, by the RADIOFOREGROUNDS grant (COMPET-05-2015, agreement number 687312) of the European Union Horizon 2020 research and innovation program, and the MIUR grant 'Finanziamento annuale individuale attivita base di ricerca'. 
Finally, MN received funding from the European Union's Horizon 2020 research and innovation programme under the Marie Sklodowska-Curie grant agreement No 707601.\\
We deeply acknowledge the CINECA award under the ISCRA initiative, for the availability of high performance computing resources and support. In particular the projects ``SIS19\_lapi", ``SIS20\_lapi" in the framework ``Convenzione triennale SISSA-CINECA".\\
The Herschel-ATLAS is a project with Herschel, which is an ESA space observatory with science instruments provided by European-led Principal Investigator consortia and with important participation from NASA. The H-ATLAS web-site is http://www.h-atlas.org. GAMA is a joint European-Australasian project based around a spectroscopic campaign using the Anglo-Australian Telescope. The GAMA input catalogue is based on data taken from the Sloan Digital Sky Survey and the UKIRT Infrared Deep Sky Survey. Complementary imaging of the GAMA regions is being obtained by a number of independent survey programs including GALEX MIS, VST KIDS, VISTA VIKING, WISE, Herschel-ATLAS, GMRT and ASKAP providing UV to radio coverage. GAMA is funded by the STFC (UK), the ARC (Australia), the AAO, and the participating institutions. The GAMA web-site is: http://www.gama-survey.org/.\\
In this work, we made extensive use of \texttt{GetDist} \citep{GETDIST}, a Python package for analysing and plotting MC samples. In addition, this research has made use of the python packages \texttt{ipython} \citep{ipython}, \texttt{matplotlib} \citep{matplotlib} and \texttt{Scipy} \citep{scipy}

\end{acknowledgements}

%
%

  \bibliographystyle{aa} 
  \bibliography{./XCORR_COSMO} 

\appendix
\section{Cross-correlation function sensitivity to the model's parameters}
\label{appendix:sens_plots}
\begin{figure*}[ht]
    \centering
    \includegraphics[width=9cm]{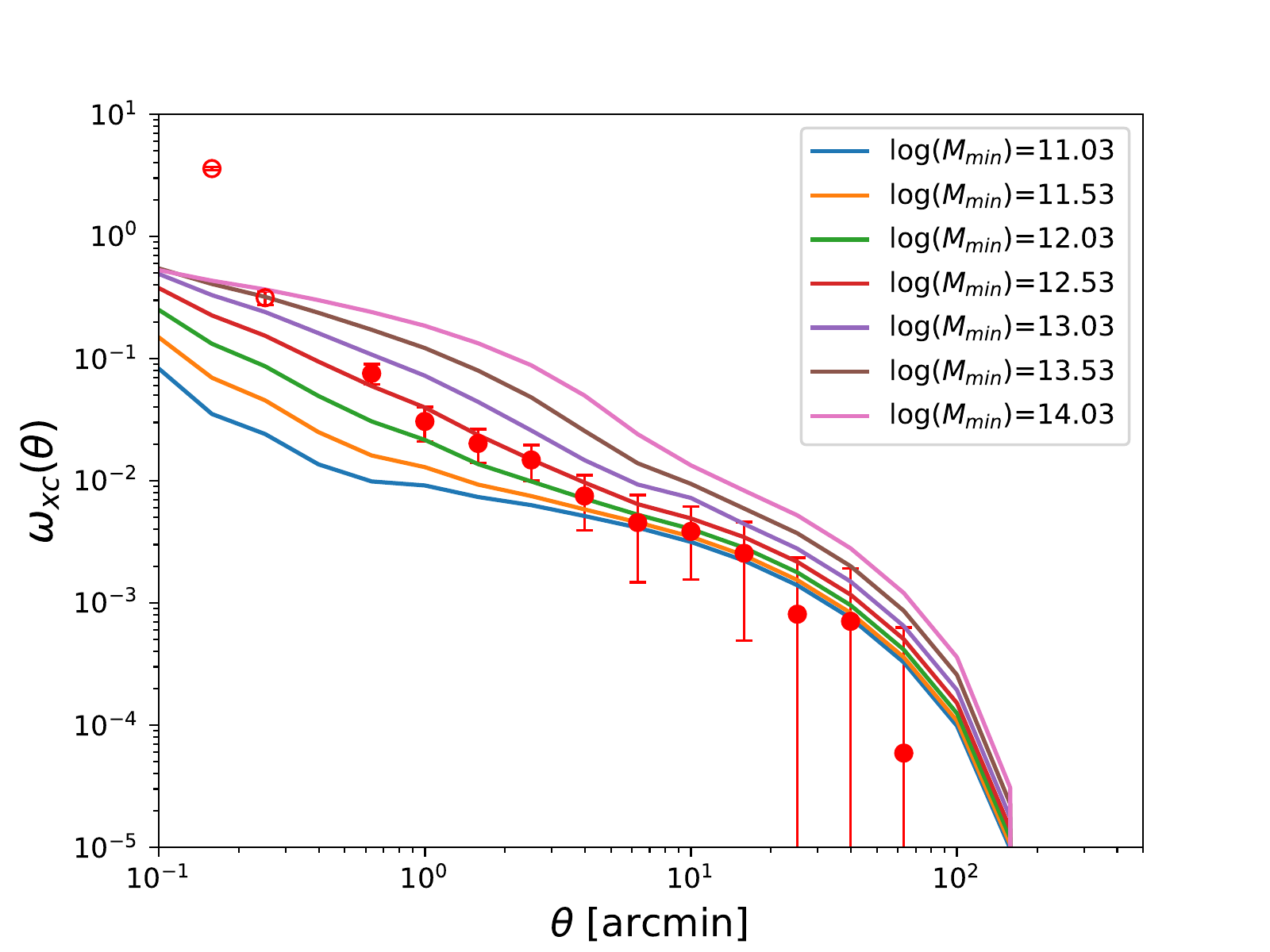}
    \includegraphics[width=9cm]{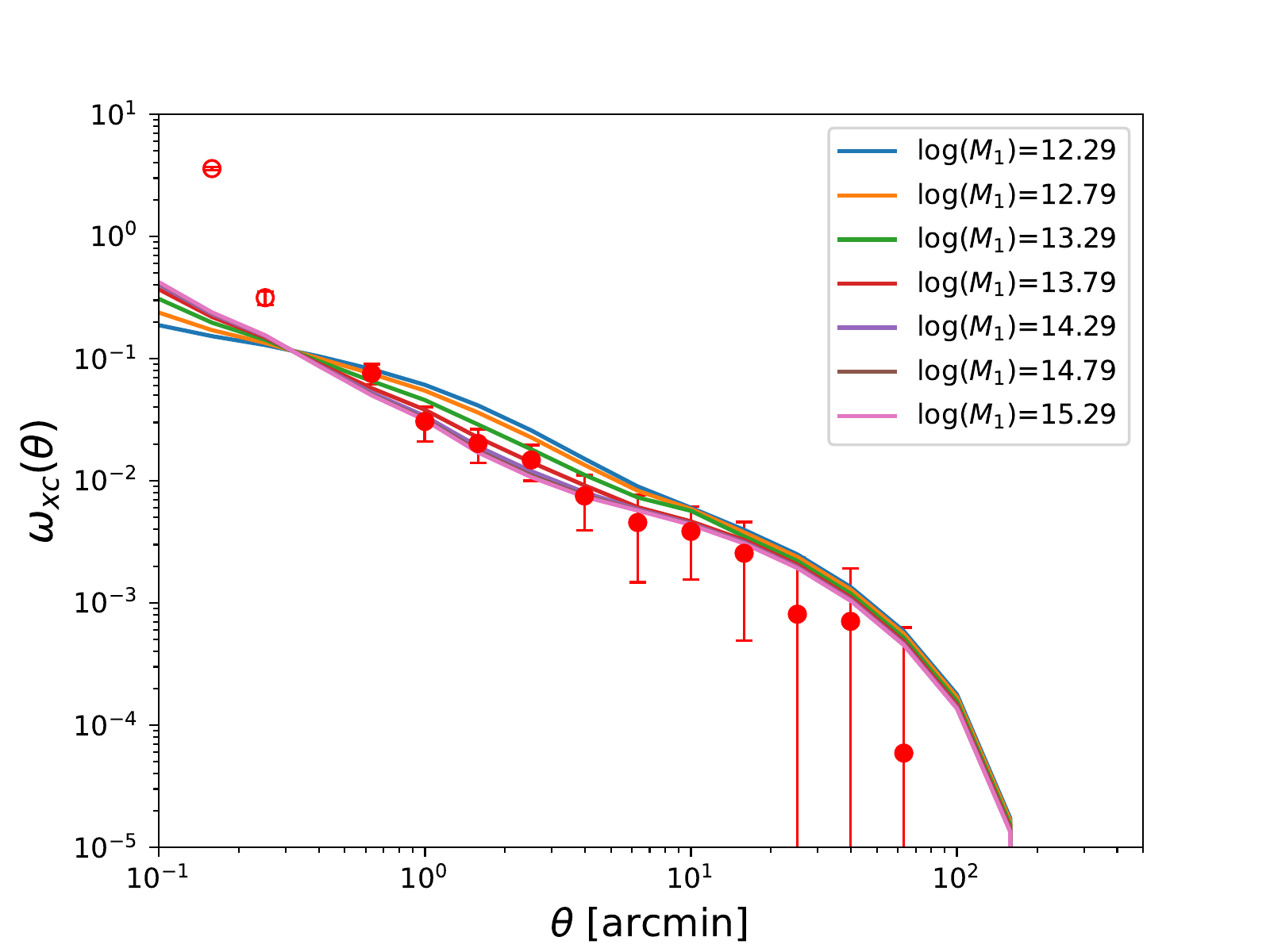}\\
    \includegraphics[width=9cm]{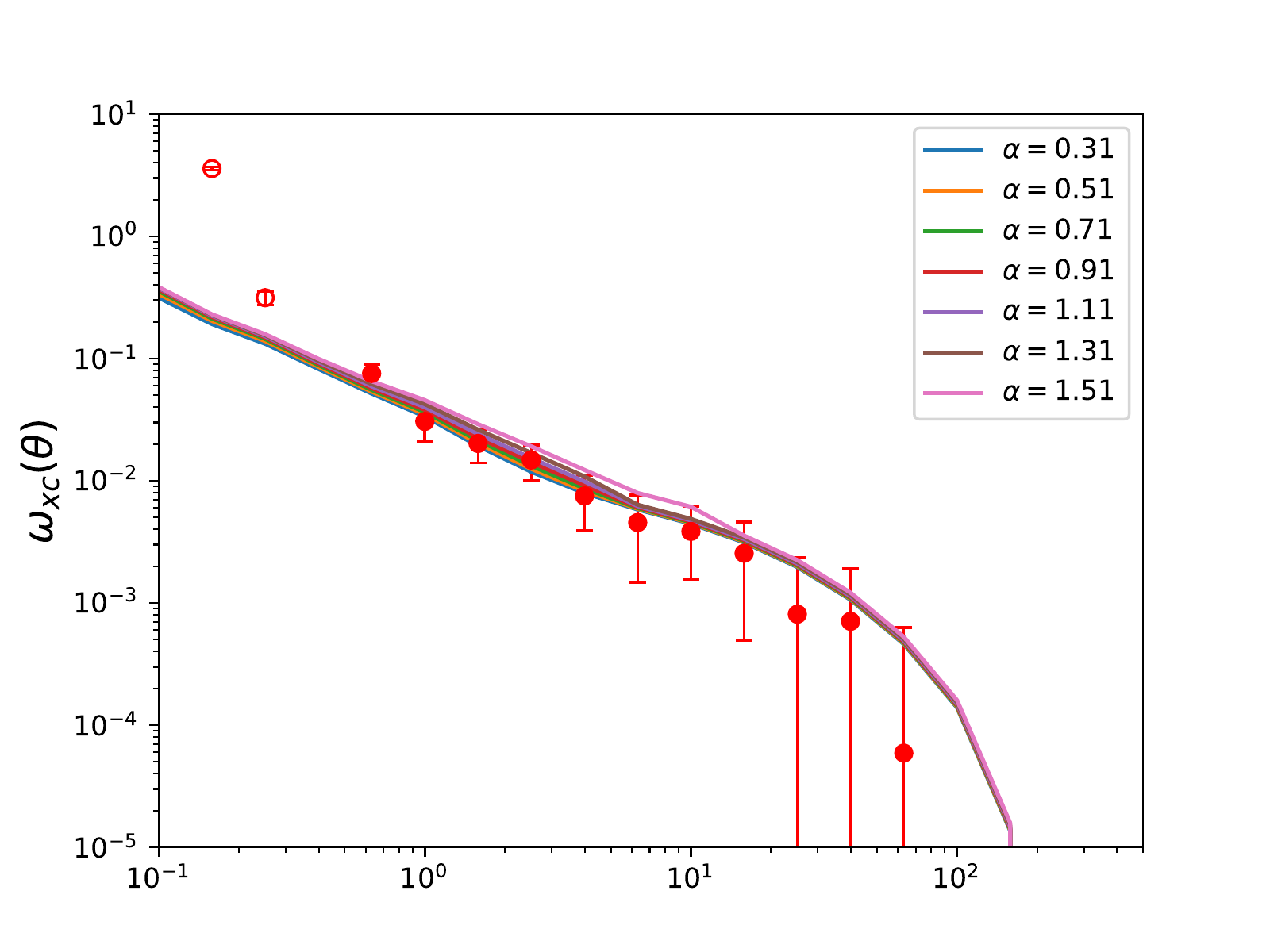}
    \includegraphics[width=9cm]{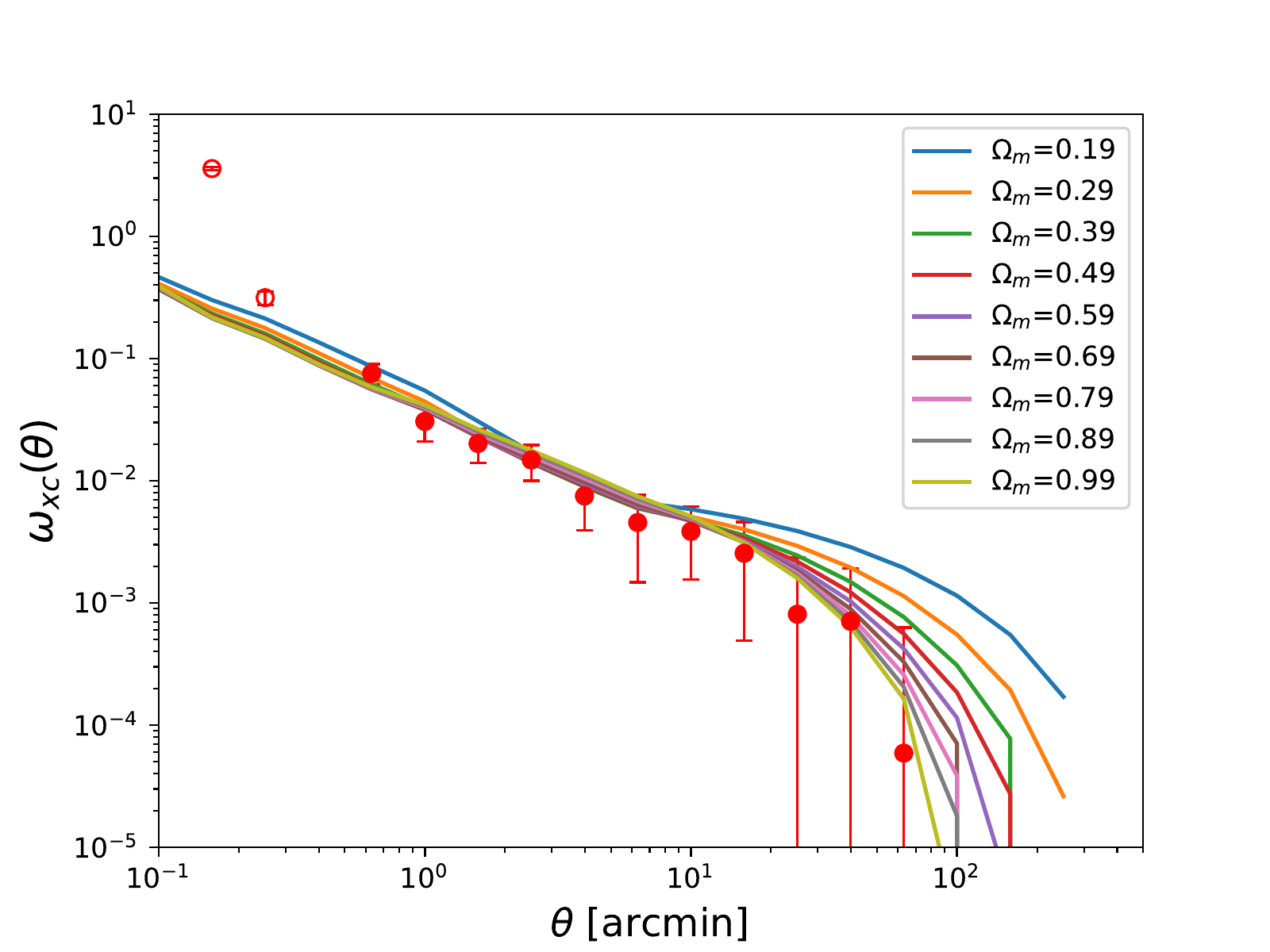}\\
    \includegraphics[width=9cm]{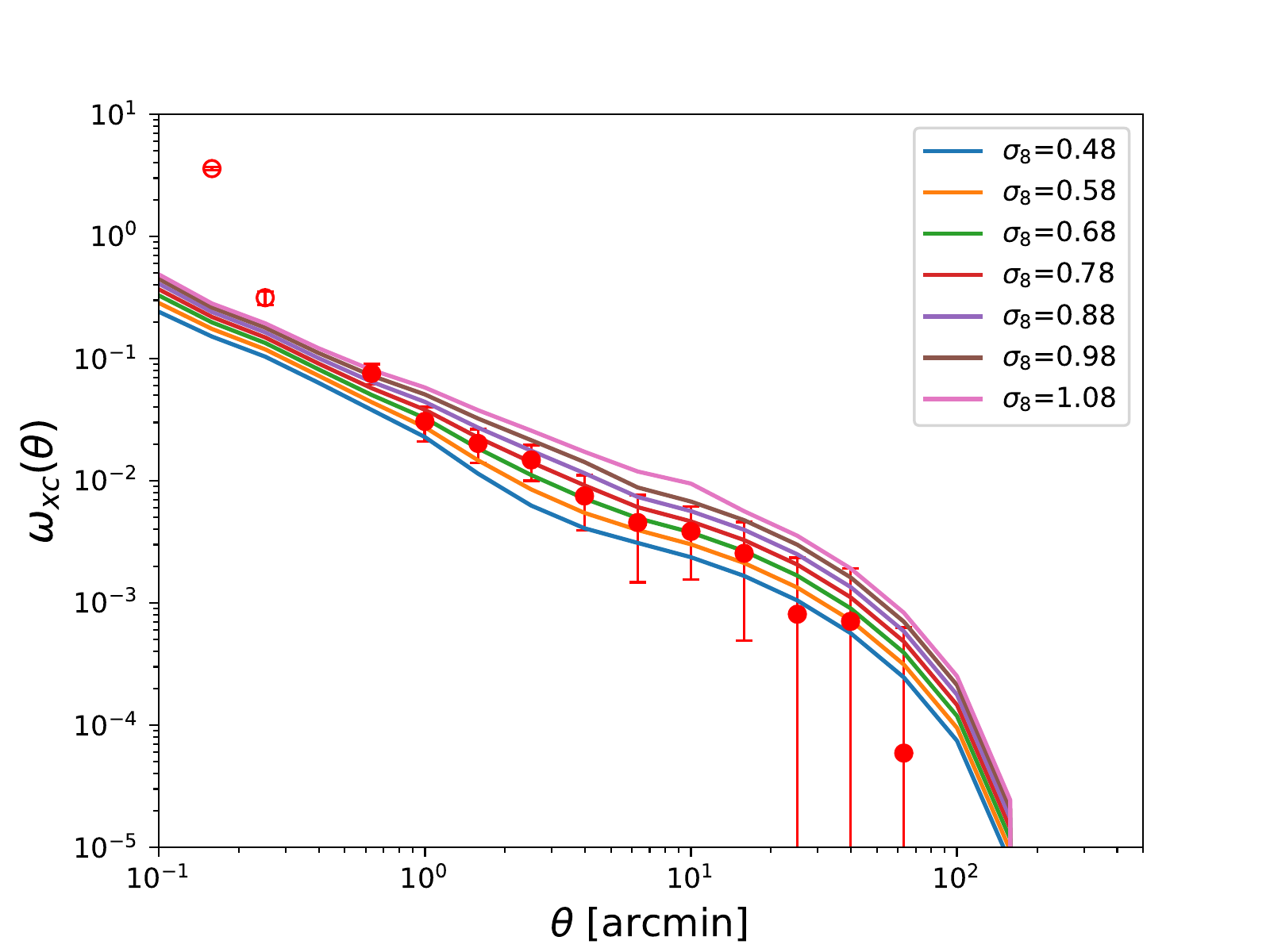}
    \includegraphics[width=9cm]{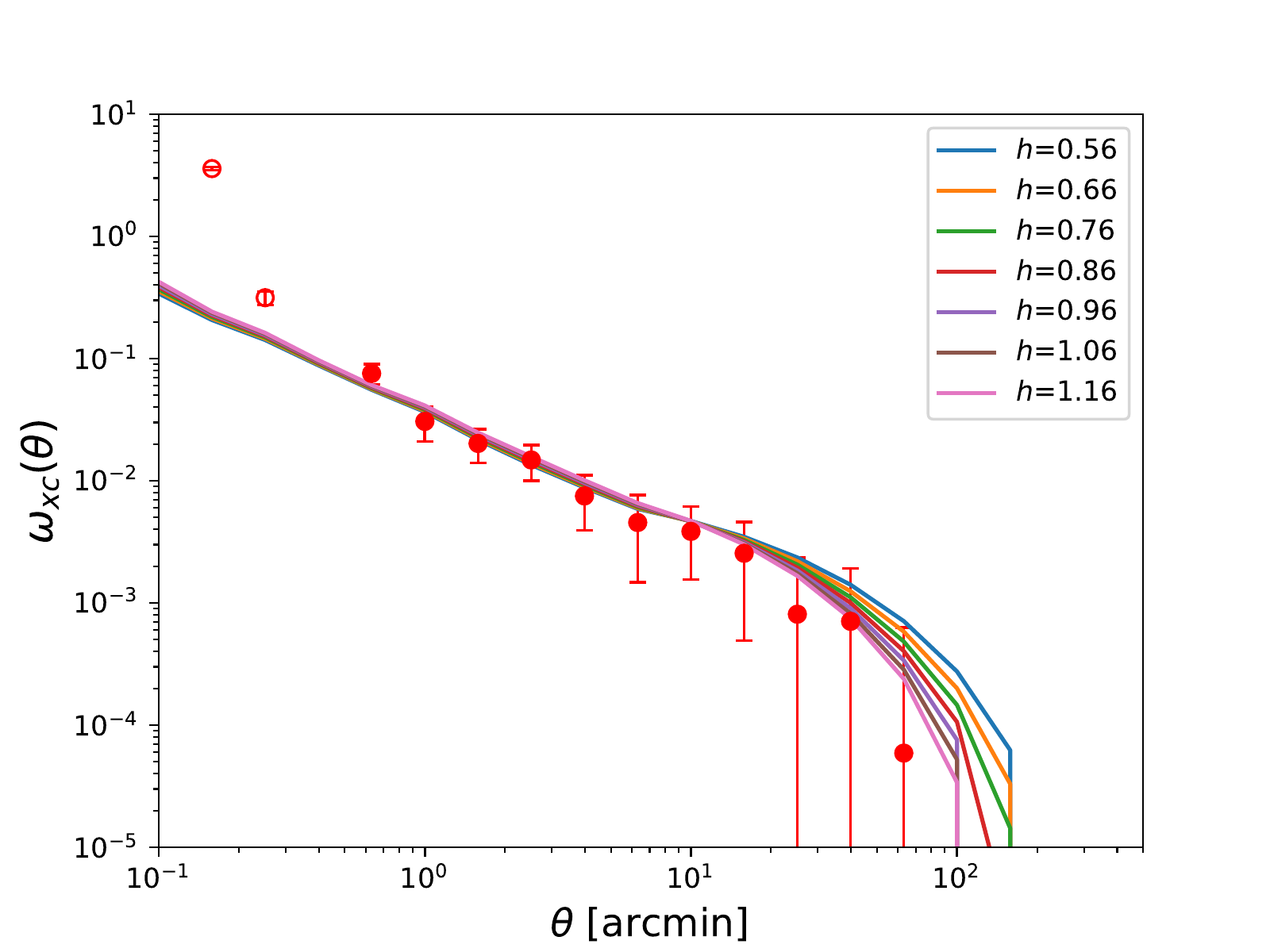}\\    
    \caption{Dependency of the cross-correlation function with each of the model's parameters (maintaining the others fixed).}
        \label{Fig:sensitivity}
\end{figure*}

Figure \ref{Fig:sensitivity} shows the dependence of the cross-correlation function with each  of the model's parameters: $M_{min}$, $M_1$, $\alpha$, $\Omega_m$, $\sigma_8$, and $H_0$ (from left to right and top to bottom). 

Of the astrophysical parameters, the most sensitive one is $M_{min}$. Considering the value ranges used in this work for $M_1$ and $\alpha$, it is clear that the variation is located only at intermediate angular scales and it is small compared with the observational uncertainties. This explains that both parameters are almost unconstrained with the cross-correlation data alone.

With respect to the cosmological parameters, the most sensitive one is $\sigma_8$. Its effect on the cross-correlation function are most important on intermediate angular scales, but it also has   a non-negligible effect on large angular scales. On the contrary, the effect of $\Omega_m$  is only important on large angular scales and   negligible   on lower angular scales. Moreover, higher values of $\sigma_8$ increase the cross-correlation signal, while the opposite occurs for $\Omega_m$. This explains how the usual $\Omega_m$-$\sigma_8$ degeneracy could be broken with the cross-correlation function once more accurate observation at large angular scales are available.

Finally, the effect produced by $H_0$ is relevant only on very large angular scales (even larger than in the $\Omega_m$ case). As currently these angular scales are those observed with the largest uncertainties, $H_0$ is unconstrained.

\end{document}